\documentclass[a4paper,11pt]{article}
\usepackage[english]{babel}
\usepackage{jheppub}
\pdfoutput=1
\usepackage[T1]{fontenc}
\usepackage{bbold}
\usepackage{float}
\usepackage{amsmath}
\usepackage{empheq}
\usepackage{booktabs}
\usepackage{color}
\usepackage[utf8]{inputenc}
\usepackage{xspace}
\usepackage{scalerel}
\usepackage[most]{tcolorbox}
\usepackage{multirow}
\usepackage{scalefnt}
\usepackage{bold-extra}
\usepackage[shortlabels]{enumitem}
\usepackage[tikz]{bclogo}
\usepackage{subcaption}
\usepackage{tikz-feynman}
\usepackage{cancel}
\usepackage{verbatim}
\usetikzlibrary{arrows,shapes}
\usepackage{afterpage}
\usepackage{graphicx,grffile}

\newcommand{\stepone}{{Step\,I}}
\newcommand{\steptwo}{{Step\,II}}
\newcommand{\stepthree}{{Step\,III}}

\usepackage[normalem]{ulem}

\newcommand\F{${\rm F}$}
\newcommand\FJ{${\rm FJ}$}
\newcommand\FJJ{${\rm FJJ}$}

\newcommand\wz{$W^\pm Z$}
\newcommand\wpz{$W^+ Z$}
\newcommand\wmz{$W^- Z$}

\newcommand{\order}[1]{{\cal O}\left(#1\right)}
\newcommand{\as}{\alpha_s}

\newcommand{\pt}{{p_{\text{\scalefont{0.77}T}}}}

\newcommand{\ptrad}{{p_{\text{\scalefont{0.77}T,rad}}}}

\newcommand{\ptz}{{p_{\text{\scalefont{0.77}T,$Z$}}}}
\newcommand{\ptw}{{p_{\text{\scalefont{0.77}T,$W$}}}}
\newcommand{\ptnu}{{p_{\text{\scalefont{0.77}T,$\nu$}}}}
\newcommand{\mtwz}{{m_{\text{\scalefont{0.77}T,$WZ$}}}}
\newcommand{\dphiwz}{{\Delta\phi_{\text{\scalefont{0.77}$WZ$}}}}
\newcommand{\dyZlW}{{|y_{\text{\scalefont{0.77}$Z$}}-y_{\text{\scalefont{0.77}$
\ell_W$}}|}}

\newcommand{\ptlnu}{{p_{\text{\scalefont{0.77}T,$\mu\nu_\mu$}}}}

\newcommand{\ptlone}{{p_{\text{\scalefont{0.77}T,$\ell_1$}}}}
\newcommand{\ptltwo}{{p_{\text{\scalefont{0.77}T,$\ell_2$}}}}
\newcommand{\ptmiss}{{p_{\text{\scalefont{0.77}T,miss}}}}

\newcommand{\ylone}{{y_{\text{\scalefont{0.77}$\ell_1$}}}}
\newcommand{\dphill}{{\Delta\phi_{\text{\scalefont{0.77}$\ell_1\ell_2$}}}}

\newcommand{\mz}{{m_{\text{\scalefont{0.77}Z}}}}
\newcommand{\mw}{{m_{\text{\scalefont{0.77}W}}}}

\newcommand{\mlnu}{{m_{\text{\scalefont{0.77}$\mu\nu_\mu$}}}}
\newcommand{\etal}{{\eta_{\text{\scalefont{0.77}$\ell$}}}}

\newcommand{\lw}{\ensuremath{\mu}}

\newcommand{\lz}{\ensuremath{e}}

\newcommand{\ptlw}{\ensuremath{p_{\text{\scalefont{0.77}T,\lw}}}}

\newcommand{\mlll}{\ensuremath{m_{\text{\scalefont{0.77}$3\ell$}}}}

\newcommand{\mtw}{\ensuremath{m_{\text{\scalefont{0.77}T,W}}}}

\newcommand{\mll}{{m_{\text{\scalefont{0.77}\lz\lz}}}}
\newcommand{\yll}{{y_{\text{\scalefont{0.77}\lz\lz}}}}

\newcommand{\qcdfull}{\ensuremath{\text{NNLO}_{\rm QCD}^{{\rm (QCD,QED)}_{\rm PS}}}}
\newcommand{\qcdqcd}{\ensuremath{\text{NNLO}_{\rm QCD}^{{\rm (QCD)}_{\rm PS}}}}

\newcommand{\addfull}{\ensuremath{\text{NNLO}_{\rm QCD}^{\rm (QCD,QED)_{\rm PS}} + \delta{\rm NLO}_{\rm EW}^{\rm (QCD,QED)_{\rm PS}}}}
\newcommand{\addqcdfull}{\ensuremath{\text{NNLO}_{\rm QCD}^{\rm (QCD,QED)_{\rm PS}} + \delta{\rm NLO}_{\rm EW}^{\rm (QED)_{\rm PS}}}}
\newcommand{\addqedfull}{\ensuremath{\text{NLO}_{\rm EW}^{\rm (QCD,QED)_{\rm PS}} + \delta{\rm NNLO}_{\rm QCD}^{\rm (QCD)_{\rm PS}}}}

\newcommand{\multfull}{\ensuremath{\text{NNLO}_{\rm QCD}^{\rm (QCD,QED)_{\rm PS}} \times \text{K-NLO}_{\rm EW}^{\rm (QCD,QED)_{\rm PS}}}}
\newcommand{\multqcdfull}{\ensuremath{\text{NNLO}_{\rm QCD}^{\rm (QCD,QED)_{\rm PS}} \times \text{K-NLO}_{\rm EW}^{\rm (QED)_{\rm PS}}}}
\newcommand{\multqedfull}{\ensuremath{\text{NLO}_{\rm EW}^{\rm (QCD,QED)_{\rm PS}} \times \text{K-NNLO}_{\rm QCD}^{\rm (QCD)_{\rm PS}}}}

\newcommand{\QCDpEW}{\ensuremath{ \text{NNLO}_{\rm QCD+EW}^{\rm (QCD, QED)_{\rm PS}}}} 
\newcommand{\QCDtEW}{\ensuremath{ \text{NNLO}_{\rm QCDxEW}^{\rm (QCD, QED)_{\rm PS}}}} 
\newcommand{\QCDtEWfo}{\ensuremath{ \text{NNLO}_{\rm QCD}^{\rm (QCD)_{\rm PS}} \times \text{K-NLO}_{\rm EW}^{\rm (f.o.)}}}

\newcommand{\muF}{{\mu_{\text{\scalefont{0.77}F}}}}
\newcommand{\muR}{{\mu_{\text{\scalefont{0.77}R}}}}

\newcommand{\Q}{{Q_{\text{\scalefont{0.77}$0$}}}}

\newcommand{\noun}[1]{{\scshape #1}}

\newcommand{\POWHEG}{\noun{Powheg}}

\newcommand{\POWHEGBOX}{\noun{Powheg-Box}}
\newcommand{\POWHEGBOXRES}{\noun{Powheg-Box-Res}}

\newcommand{\minnlo}{{\noun{MiNNLO$_{\rm PS}$}}}
\newcommand{\Matrix}{{\noun{Matrix}}}
\newcommand{\OpenLoops}{{\noun{OpenLoops}}}
\newcommand{\PYTHIA}[1]{\noun{Pythia{#1}}}

\newcommand{\setupinclusive}{{\tt inclusive setup}}
\newcommand{\setupfiducial}{{\tt fiducial setup}}

\newcommand{\citere}[1]{ref.\,\cite{#1}}

\newcommand{\citeres}[1]{refs.\,\cite{#1}}

\newcommand{\eqn}[1]{eq.\,(\ref{#1})}

\newcommand{\fig}[1]{figure\,\ref{#1}}
\newcommand{\Fig}[1]{Figure\,\ref{#1}}
\newcommand{\figs}[1]{figures\,\ref{#1}}
\newcommand{\tab}[1]{table\,\ref{#1}}
\newcommand{\sct}[1]{section~\ref{#1}}

\newcommand{\LambdaPWG}{\Lambda_{\rm pwg}}

\usepackage{xcolor}

\newcommand{\tmop}[1]{\ensuremath{\operatorname{#1}}}

\newtcolorbox{empheqboxed}{colback=white!35, 
 colframe=black,
 width=\textwidth,
 sharpish corners,
 top=-2mm,
 bottom=0pt
}

\title{\boldmath{$W^\pm Z$} production at NNLO QCD and NLO EW matched to parton showers with M{\scalefont{0.77}I}NNLO\boldmath{$_{\text{PS}}$}}

\author[]{Jonas M. Lindert,$^{a}$}
\author[]{Daniele Lombardi,$^{b}$}
\author[]{Marius Wiesemann,$^{b}$}
\author[]{Giulia Zanderighi$^{b,c}$}
\author[]{\\ and Silvia Zanoli$^{b}$}

\emailAdd{lindert@sussex.ac.uk}
\emailAdd{lombardi@mpp.mpg.de}
\emailAdd{wieseman@mpp.mpg.de}
\emailAdd{zanderi@mpp.mpg.de}
\emailAdd{zanoli@mpp.mpg.de}

\affiliation[]{$^{a}$Department of Physics and Astronomy, University of Sussex, Brighton BN1 9QH, UK}
\affiliation[]{$^{b}$Max-Planck-Institut f\"ur Physik, F\"ohringer Ring 6,
  80805 M\"unchen, Germany}
\affiliation[]{$^{c}$Physik-Department, Technische Universit\"at M\"unchen, James-Franck-Strasse 1, 85748 Garching, Germany}

\abstract{We consider $W^\pm Z$ production in hadronic collisions and
  present high-precision predictions in QCD and electroweak (EW)
  perturbation theory matched to parton showers. To this end, we match
  next-to-next-to-leading order QCD corrections to parton showers
  using the \minnlo{} method and consistently combine them with
  next-to-leading order EW corrections matched to parton showers.
  This is the first time such accuracy in the event generation is
  achieved for any collider process, and we study in detail the impact
  of different choices in the combination of QCD and EW corrections as
  well as QCD and QED showers.  Spin correlations, interferences and
  off-shell effects are retained by considering the full leptonic
  processes $pp \to \ell^+\ell^-\ell'^\pm \nu_\ell'$ with
  $\ell'\neq\ell$ and $\ell'=\ell$ without approximations, and the
  matching to QED radiation is performed preserving the resonance
  structure of the process.  We find that NNLO QCD predictions
  including QCD and QED shower effects provide a very good
  approximation in the bulk-region of the phase space, while EW
  effects become increasingly important in the high-energy tails of
  kinematic distributions.  Our default predictions are in excellent
  agreement with recent ATLAS data.}


\preprint{MPP-2022-98}

\begin{document}

\maketitle

\section{Introduction}
\label{sec:intro}

Among the major theoretical challenges for todays physics program at
CERN's Large Hadron Collider (LHC) are precision simulations of
proton--proton reactions based on calculations in QCD and EW
perturbation theory at the highest possible order.  The high demand
for such accurate predictions is the result of the remarkable
performance of the LHC experiments, which keep decreasing the
experimental uncertainties of various inclusive and differential
cross-section measurements. Moreover, without clear hints for new
physics at the LHC thus far, data--theory comparisons at high
precision have become a promising path towards the observation of
deviations from the Standard Model (SM) picture.

The class of processes where a pair of vector bosons (decaying to
leptons) is produced represents a highly relevant set of LHC reactions
in these endeavours. Not only do these processes provide direct access
to trilinear gauge couplings, which are often modified or added as new
contributions with respect to the SM Lagrangian in various
beyond-the-SM (BSM) theories, they also yield a central test of the
gauge-symmetry structure of EW interactions within the SM, as any
small deviation from the expected rates or shapes of distributions
could be a signal of new physics. In this context, \wz{} production is
particularly interesting due its relatively large cross section and
clean experimental signature, which allows very accurate experimental
measurements of this process at the LHC. Moreover, \wz{} production
holds a special place in BSM searches, both as signal and as
background. Indeed, even within the SM \wz{} production features a
trilinear gauge coupling that enters already at tree-level. On the
other hand, the \wz{} process yields an important SM background in
many BSM resonance searches, such as for supersymmetric particles (see
e.g.\ \citere{Morrissey:2009tf}).

Measurements of the \wz{} cross section have been performed both at
the Tevatron~\cite{Aaltonen:2012vu,Abazov:2012cj} and at the LHC for
centre-of-mass energies of
7\,TeV~\cite{Aad:2012twa,Khachatryan:2016poo},
8\,TeV~\cite{Aad:2016ett,Khachatryan:2016poo} and
13\,TeV~\cite{Aaboud:2016yus,Khachatryan:2016tgp,ATLAS:2019bsc,CMS:2021icx}.
On the theory side, next-to-leading-order (NLO) predictions in QCD for
\wz{} production were obtained long ago
\cite{Ohnemus:1991gb,Ohnemus:1994ff,Campbell:1999ah,Dixon:1999di,Campbell:2011bn}. Corresponding
results for polarized \wz{} production became available only recently
in the double-pole approximation~\cite{Denner:2020eck}.  The
computation of the \wz{}+${\rm jet}$ cross section at NLO QCD was
presented in \citere{Campanario:2010hp}.  The first next-to-NLO (NNLO)
QCD accurate predictions were obtained for the inclusive \wz{} cross
section in~\citere{Grazzini:2016swo}, which were later extended to the
fully differential predictions including leptonic decays in
\citere{Grazzini:2017ckn}. The computation of NNLO QCD corrections are
publicly available in the parton-level Monte Carlo framework
\Matrix~\cite{Grazzini:2017mhc} and MCFM~\cite{Campbell:2022gdq}. In
the \Matrix{} framework also the effect of the $b$-space resummation
of large logarithmic terms at small transverse momenta of the \wz{}
system up to next-to-next-to-leading logarithmic accuracy (NNLL) has
been incorporated~\cite{Grazzini:2015wpa}. More recently, the {\sc
  Matrix+RadISH} framework was introduced
\cite{Kallweit:2020gva,Wiesemann:2020gbm,Grazzini:2017mhc,Monni:2016ktx,Bizon:2017rah,Monni:2019yyr},
which makes NNLO+N$^3$LL predictions for the \wz{} transverse
momentum, NNLO+NNLL predictions for the transverse momentum of the
leading jet, as well as their joint resummation at NNLO+NNLL publicly
available.  NLO EW corrections to \wz{} production are known for both
on-shell \wz{} production \cite{Bierweiler:2013dja,Baglio:2013toa} and
including off-shell leptonic decays~\cite{Biedermann:2017oae}.  The
NLO EW predictions have been combined with NNLO QCD corrections and
are publicly available as provided by
\Matrix+\OpenLoops~\cite{Grazzini:2019jkl}.

However, apart from idealised parton-level perturbative calculations
at higher orders, full-fledged Monte-Carlo simulations that include
higher-order corrections are becoming more and more important, as they
pair a realistic modelling of LHC events, including effects from QCD
and QED parton showers, hadronization and multiple-parton
interactions, with higher-order perturbative information. To this end,
the inclusion of NNLO QCD corrections in parton-shower simulations
(NNLO+PS) has been a very active research topic in the past ten years,
which has led to the formulation of various approaches
\cite{Hamilton:2012rf,Alioli:2013hqa,Hoeche:2014aia,Monni:2019whf,Monni:2020nks}
and ultimately to a remarkable progress in NNLO+PS calculations for a
number of colour-singlet production processes, including
$H$~\cite{Hamilton:2013fea,Hoche:2014dla,Monni:2019whf,Monni:2020nks},
$Z$/$W^\pm$~\cite{Hoeche:2014aia,Karlberg:2014qua,Alioli:2015toa,Monni:2019whf,Monni:2020nks,Alioli:2021qbf},
$ZH$/$W^\pm
H$~\cite{Astill:2016hpa,Astill:2018ivh,Alioli:2019qzz,Zanoli:2021iyp,Haisch:2022nwz},
$\gamma\gamma$~\cite{Alioli:2020qrd,Gavardi:2022ixt},
$Z\gamma$~\cite{Lombardi:2020wju,Lombardi:2021wug},
$ZZ$~\cite{Alioli:2021egp,Buonocore:2021fnj}, and $W^+W^-$
\cite{Re:2018vac,Lombardi:2021rvg}. With top-quark pair production
even the first production process with colour charges in the final
state has been computed at
NNLO+PS~\cite{Mazzitelli:2020jio,Mazzitelli:2021mmm}.  Recently, also
the matching of NLO EW corrections to QED and QCD parton showers has
been considered for massive diboson
processes~\cite{Chiesa:2020ttl,Bothmann:2021led}. In order to ensure a
consistent off-shell description in \citere{Chiesa:2020ttl} the
matching had to be performed in a resonance-aware fashion as provided
by the \POWHEGBOXRES{} framework~\cite{Jezo:2015aia}.

In this paper we present the first NNLO+PS calculation for \wz{}
production in QCD, and we combine these results with NLO EW
corrections matched consistently to parton showers. More precisely, we
consider the full process that leads to three leptons and one
neutrino, $pp\to \ell^{'\pm} \nu_{\ell^{'}} \ell^+\ell^-+X$, in both
the same-flavour ($\ell' = \ell$) and the different-flavour
($\ell'\neq\ell$) channel, taking into account all non-resonant,
single-resonant and double-resonant components with all interference
effects, spin correlations and off-shell effects in the complex-mass
scheme~\cite{Denner:2005fg}, and we combine NNLO QCD and NLO EW
corrections properly matched to QCD and QED parton showers.  We
validate our predictions against the corresponding fixed-order
calculations, and we study various possible schemes for the
combination of QCD and EW corrections. These different schemes not
only distinguish between additive and multiplicative combinations, but
also avoid the double-counting of QCD and QED radiation effects in
different ways, thereby differing by all-order terms that are beyond
accuracy.
We note that among all diboson processes \wz{} production offers a
valuable application to perform this study. In particular, unlike
colour-neutral diboson production, the \wz{} process does not involve
a loop-induced gluon fusion channel, which receives sizeable
higher-order QCD corrections that are crucial to obtain accurate
predictions~\cite{Caola:2015psa,Caola:2016trd,Alioli:2016xab,Grazzini:2018owa,Grazzini:2020stb,Grazzini:2021iae,Alioli:2021wpn}. \wz{}
production does also not feature photon-induced subprocesses at the
Born level.  Our calculation is performed and implemented within
\POWHEGBOXRES{} and will be made publicly available.

This manuscript is organized as follows: in \sct{sec:calculation} we
describe our calculation including the relevant information on the
process (\sct{sec:process}), on the \minnlo{} method and its practical
implementation to obtain a \wz{} NNLO+PS QCD generator
(\sct{sec:minnlo}), on the NLO+PS EW implementation
(\sct{sec:nloewps}), on constraining QCD and QED shower radiation that
is necessary to preserve the accuracy of the predictions
(\sct{sec:veto}) and on the combination of QCD and EW corrections
(\sct{sec:combination}). In \sct{sec:results} phenomenological results
are presented, where we first discuss our input settings
(\sct{sec:settings}) and subsequently validate our NNLO QCD and NLO EW
accurate event simulations against fixed-order results
(\sct{sec:validation}), compare different QCD and EW combination
schemes (\sct{sec:comparisonmatching}), and finally present a
comparison of our best predictions to recent ATLAS data
(\sct{sec:comparisondata}).  We conclude in \sct{sec:summary}.

\section{Outline of the calculation}
\label{sec:calculation}

\subsection{Description of the process and notation}
\label{sec:process}

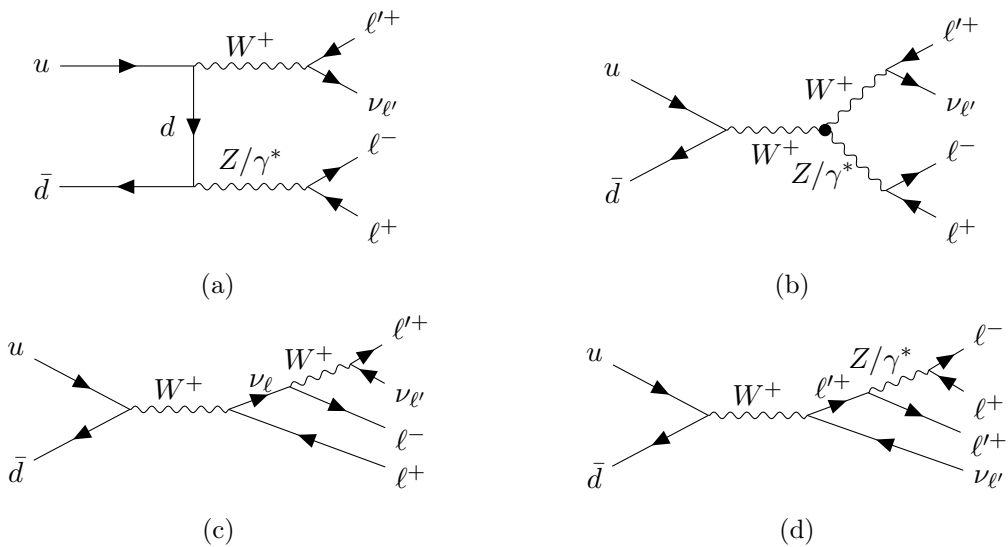
\begin{figure}[b]
  \begin{center}
    \begin{subfigure}[b]{0.5\linewidth}
      \centering
      \begin{tikzpicture}
        \begin{feynman}
          \vertex (a1) {\(u\)};
          \vertex[below=1.6cm of a1] (a2){\(\bar d\)};
          \vertex[right=2cm of a1] (a3);
          \vertex[right=2cm of a2] (a4);
          \vertex[right=1.5cm of a3] (a5);
          \vertex[right=1cm of a5] (a20);
          \vertex[above=0.3cm of a20] (a7){\(\ell'^+\)} ;
          \vertex[below=0.3cm of a20] (a8){\(\nu_{\ell'}\)};
          \vertex[right=1.5cm of a4] (a6);
          \vertex[right=1cm of a6] (a21);
          \vertex[above=0.3cm of a21] (a9){\(\ell^-\)} ;
          \vertex[below=0.3cm of a21] (a10){\(\ell^+\)};

          \diagram* {
            {[edges=fermion]
              (a1)--(a3)--[edge label'=\(d\;\)](a4)--(a2),
              (a7)--(a5)--(a8),
              (a10)--(a6)--(a9),
            },
            (a3) -- [boson, edge label=\(W^+\)] (a5),
            (a4) -- [boson, edge label=\(Z/\gamma^*\)] (a6),
          };

        \end{feynman}
      \end{tikzpicture}
      \caption{}
        \label{subfig:res1}
    \end{subfigure}%
\begin{subfigure}[b]{.5\linewidth}
\centering
  \begin{tikzpicture}
    \begin{feynman}
      \vertex (a1) {\(u\)};
      \vertex[below=1.6cm of a1] (a2){\(\bar d\)};
      \vertex[below=0.8cm of a1] (a3);
      \vertex[right=1.5cm of a3] (a4);
      \vertex[right=1.3cm of a4] (a5);
      \vertex[dot,fill=black] (d) at (a5){};
      \vertex[right=0.8cm of a5] (a6);
      \vertex[below=0.8cm of a6](a8);
      \vertex[above=0.8cm of a6](a7);

      \vertex[right=1cm of a7] (a9);
      \vertex[above=0.25cm of a9] (a11){\(\ell'^+\)} ;
      \vertex[below=0.25cm of a9] (a12){\(\nu_{\ell'}\)};

      \vertex[right=1cm of a8] (a10);
      \vertex[above=0.25cm of a10] (a13){\(\ell^-\)} ;
      \vertex[below=0.25cm of a10] (a14){\(\ell^+\)};
      \diagram* {
        {[edges=fermion]
          (a1)--(a4)--(a2),
          (a11)--(a7)--(a12),
          (a14)--(a8)--(a13),
        },
        (a4) -- [boson, edge label'=\(W^+\)] (a5),
        (a5) -- [boson,edge label=\(W^+\),inner sep=0pt] (a7),
        (a5) -- [boson,edge label'=\(Z/\gamma^*\), inner sep = 0pt] (a8),
      };
      
    \end{feynman}
  \end{tikzpicture}
  \caption{}
        \label{subfig:res2}
\end{subfigure}%

\begin{subfigure}[b]{0.5\linewidth}
      \centering
\begin{tikzpicture}
  \begin{feynman}
    \vertex (a1) {\(u\)};
    \vertex[below=1.6cm of a1] (a2){\(\bar d\)};
    \vertex[below=0.8cm of a1] (a3);
    \vertex[right=1.5cm of a3] (a4);

    \vertex[right=1.3cm of a4] (a5);

    \vertex[right=0.8cm of a5](a6);
    \vertex[above=0.3cm of a6](a7);

    \vertex[right=0.8cm of a7](a8);
    \vertex[above=0.3cm of a8](a9);

    \vertex[right=0.8cm of a9](a10);
    \vertex[above=0.2cm of a10](a11){\(\ell'^+\)};
    \vertex[below=0.2cm of a10](a12){\({\nu}_{\ell'}\)};
    \vertex[below=0.7cm of a10](a13){\(\ell^-\)};
    \vertex[below=1.2cm of a10](a14){\(\ell^+\)};

     \diagram* {
       {[edges=fermion]
         (a1)--(a4)--(a2),
         (a14)--(a5)--[edge label=\(\nu_\ell\),inner sep=0pt,near end](a7)--(a13),
         (a12)--(a9)--(a11),
       },
       (a4) -- [boson, edge label=\(W^+\)] (a5),
       (a7) -- [boson, edge label=\(W^+\),inner sep=0pt,near end] (a9),
       };

  \end{feynman}
\end{tikzpicture}
\caption{}
        \label{subfig:res3}
\end{subfigure}
\begin{subfigure}[b]{0.49\linewidth}
      \centering
\begin{tikzpicture}
  \begin{feynman}
    \vertex (a1) {\(u\)};
    \vertex[below=1.6cm of a1] (a2){\(\bar d\)};
    \vertex[below=0.8cm of a1] (a3);
    \vertex[right=1.5cm of a3] (a4);

    \vertex[right=1.3cm of a4] (a5);

    \vertex[right=0.8cm of a5](a6);
    \vertex[above=0.3cm of a6](a7);

    \vertex[right=0.8cm of a7](a8);
    \vertex[above=0.3cm of a8](a9);

    \vertex[right=0.8cm of a9](a10);
    \vertex[above=0.2cm of a10](a11){\(\ell^-\)};
    \vertex[below=0.2cm of a10](a12){\(\ell^+\)};
    \vertex[below=0.7cm of a10](a13){\(\ell'^+\)};
    \vertex[below=1.2cm of a10](a14){\(\nu_{\ell'}\)};

     \diagram* {
       {[edges=fermion]
         (a1)--(a4)--(a2),
         (a14)--(a5)--[edge label=\(\ell'^+\),inner sep=0pt,near end](a7)--(a13),
         (a12)--(a9)--(a11),
       },
       (a4) -- [boson, edge label=\(W^+\)] (a5),
       (a7) -- [boson, edge label=\(Z/\gamma^*\),inner sep=0pt,near end] (a9),
       };

  \end{feynman}
\end{tikzpicture}
\caption{}
        \label{subfig:res3}
\end{subfigure}
\end{center}
  \caption{\label{DiagramsWZ} Sample LO diagrams for (a) $t$-channel
    \wpz{} production, (b) $s$-channel \wpz{} production and (c,d)
    DY-type \wpz{} production channels. The corresponding diagrams for
    \wmz{} production can be obtained via charge conjugation.  }
\end{figure}

We consider the process
\begin{align}
pp\to \ell^{'\pm} \nu_{\ell^{'}} \ell^+\ell^-+X\,,
\label{eq:process}
\end{align}
for any combination of massless leptons $\ell,\ell^\prime
\in\{e,\mu,\tau\}$ with both different flavours $\ell\neq\ell^\prime$
and same flavours $\ell=\ell^\prime$.  All possible same-flavour and
different-flavour lepton combinations have been implemented in the
Monte Carlo generator that will be made publicly available within
\POWHEGBOXRES{} in the near future.  Resonant and non-resonant
topologies leading to this process, off-shell effects, interferences
and spin correlations are taken into account. At LO \wz{} production
is quark induced and of $\mathcal{O}(\alpha^4)$, where $\alpha$
denotes the electroweak coupling. In this counting the leptonic decays
of the vector bosons are included. Sample LO diagrams are shown in
\fig{DiagramsWZ}, including $t$-channel, $s$-channel and Drell-Yan
type contributions. While at higher orders in QCD perturbation theory
all combinations of partons in the initial state contribute to the
process, due to charge conservation no additional large
$\mathcal{O}(\alpha^4\alpha_s^2)$ contribution of loop-induced
gluon-fusion type is present for the process at hand, unlike for the
production of charge-neutral diboson systems.  However, \wz{}
production is subject to the so-called radiation zero effect at
LO~\cite{Brown:1979ux}, which is caused by the vanishing of the
leading helicity amplitudes in some kinematic configurations. This
fact renders the LO prediction of the process unreliable, as
higher-order corrections become particularly important and
large. Consequently, even the NNLO corrections are still of the order
of 10--15\% \cite{Grazzini:2016swo}.  Sample diagrams that contribute
at NLO QCD, i.e.\ $\mathcal{O}(\alpha^4\alpha_s)$, and at NNLO QCD,
i.e.\ $\mathcal{O}(\alpha^4\alpha_s^2)$, can be found in Fig.~3
of~\citere{Grazzini:2019jkl}.

As far as EW corrections are concerned, we consider contributions up
to $\mathcal{O}(\alpha^5)$. At this order virtual corrections enter
only through the $q \bar q$ channel and involve one-loop diagrams with
various combinations of $W$-bosons, $Z$-bosons, photons, Higgs bosons,
and fermions (including heavy quarks) in the loop.  In our calculation
real radiation contributions at NLO EW correspond to photon
radiation. Representative Feynman diagrams can be found in Fig.~4
of~\citere{Grazzini:2019jkl}.  Up to the perturbative order under
consideration no photon--photon induced contributions appear.  There
are in principle photon--quark induced contributions with an
additional quark in the final state. The computation of such
configurations are currently not supported by
\POWHEGBOXRES. Furthermore, since the photon flux in the proton is
suppressed by an additional relative $O(\alpha)$ times a collinear
logarithm $L$, we do not consider these photon--quark channels in our
computation as they would enter at $\mathcal{O}(\alpha^6 L)$. We note
however that, as shown in \citere{Biedermann:2017oae}, these channels
can yield significant contributions in various high-energy tails
driven by configurations where the initial-state photon couples
directly to a t-channel $W$-boson propagator. Therefore, in the future
they should be considered also in parton-shower matched
predictions. Nevertheless, as discussed in \citere{Grazzini:2019jkl}
such photon-induced channels should be combined in a purely additive
way with QCD higher-order corrections for the $q\bar q$ channels.  As
a result, those contributions can be implemented independently from
the present calculation and added to it incoherently in the
future.

Our calculation involves all contributions to NNLO in QCD and the
discussed ones to NLO EW including their subsequent matching to QCD
and QED parton showers. The NNLO QCD and NLO EW parton-shower matched
predictions are obtained independently and then combined based on
appropriate prescriptions that will be discussed in
Section~\ref{sec:combination}.  In order to simplify the notation we
introduce the labels NNLO$_{\rm QCD}$ and NLO$_{\rm EW}$ for the
fixed-order predictions, and correspondingly NNLO$_{\rm QCD}$+PS and
NLO$_{\rm EW}$+PS for their matching to the shower.  Where nothing is
specified, the parton shower includes both QCD and QED radiation. An
appropriate notation is introduced in
section~\ref{sec:comparisonmatching} when predictions combined with
only QCD or only QED showering are considered.

\subsection{\minnlo{} method and NNLO$_{\rm QCD}$+PS implementation}
\label{sec:minnlo}

Our implementation of the NNLO$_{\rm QCD}$+PS generator for \wz{}
production relies on the \minnlo{} method. \minnlo{} was originally
formulated and applied to $2\to 1$
processes~\cite{Monni:2019whf,Monni:2020nks} and later extended to
generic colour-singlet processes \cite{Lombardi:2020wju}, and to
heavy-quark pair production
\cite{Mazzitelli:2020jio,Mazzitelli:2021mmm}.  We refer to the
respective publications for a detailed description of the method, and
here instead sketch the main ideas and salient features in a
simplified notation.

The matching of NNLO corrections with a parton shower for a system
\F{} of colour-singlet particles in the \minnlo{} method proceeds in
three steps: First, in \stepone{}, \F{} in association with one light
parton is generated at NLO inclusively over the second radiation via
\POWHEG{}~\cite{Nason:2004rx,Nason:2006hfa,Frixione:2007vw,Alioli:2010xd}.
Second, \steptwo{} includes higher-order corrections and an
appropriate Sudakov form factor such that the cross section remains
finite in the limit where the light partons become unresolved, and the
simulation is rendered NNLO accurate for inclusive \F{} production.
Third, in \stepthree{} the second radiated parton is generated
exclusively through \POWHEG{} (accounted for inclusively in
\stepone{}) and subsequent emissions are generated through the
appropriately restricted parton shower.  Given that the emissions are
correctly ordered in $p_T$ (when using $p_T$-ordered showers) and the
Sudakov form factor in \steptwo{} matches the leading logarithms
resummed by the parton shower, \minnlo{} preserves the (leading
logarithmic) accuracy of the parton shower.

Symbolically, the fully differential \minnlo{} cross section can be
written as a \POWHEG{} calculation for \F{} plus one light parton
(\FJ{}), while including NNLO accuracy for \F{} production through a
modification of the standard \POWHEG{} $\bar B$ function:
\begin{align}
\label{eq:minnlo}
      {\rm d}\sigma_{\rm\scriptscriptstyle F}^{\rm MiNNLO_{PS}}={\rm
        d}\Phi_{\scriptscriptstyle\rm FJ}\,\bar{B}^{\,\rm MiNNLO_{\rm
          PS}}\,\times\,\left\{\Delta_{\rm pwg}(\Lambda_{\rm pwg})+
      {\rm d}\Phi_{\rm rad}\Delta_{\rm
        pwg}(\ptrad)\,\frac{R_{\scriptscriptstyle\rm
          FJ}}{B_{\scriptscriptstyle\rm FJ}}\right\}\,,
\end{align}
where the modified $\bar B$ function $\bar{B}^{\,\rm MiNNLO_{\rm PS}}$
reads
  \begin{align}
\label{eq:bbar}
\bar{B}^{\,\rm MiNNLO_{\rm PS}}= e^{-S}\,\left\{\frac{{\rm
    d}\sigma^{(1)}_{\scriptscriptstyle\rm FJ}}{{\rm
    d}\Phi_{\scriptscriptstyle\rm FJ}}\big(1+S^{(1)}\big)+\frac{{\rm
    d}\sigma^{(2)}_{\scriptscriptstyle\rm FJ}}{{\rm
    d}\Phi_{\scriptscriptstyle\rm
    FJ}}+\left(D-D^{(1)}-D^{(2)}\right)\times F^{\rm corr}\right\}\,.
  \end{align}
In~\eqn{eq:minnlo} $\Delta_{\rm pwg}$ denotes the \POWHEG{} Sudakov
form factor, having a default cutoff of $\LambdaPWG=0.89$\,GeV, while
$\Phi_{\tmop{rad}}$ and $\ptrad$ are the phase space and the
transverse momentum of the second radiation (i.e.\ the real radiation
with respect to \FJ{}), respectively. The squared tree-level matrix
elements for \FJ{} production and \FJJ{} production are
$B_{\scriptscriptstyle\rm FJ}$ and $R_{\scriptscriptstyle\rm FJ}$, and
$\Phi_{\scriptscriptstyle\rm FJ}$ indicates the \FJ{} phase
space. In~\eqn{eq:bbar} ${\rm d}\sigma^{(1,2)}_{\scriptscriptstyle\rm
  FJ}$ denote the LO and NLO differential \FJ{} cross sections and
$e^{-S}$ is the Sudakov form factor in the \F{} transverse momentum
($\pt$), with $S^{(1)}$ being the $\mathcal{O}(\as)$ term in the
expansion of its exponent.  The last term of the $\bar{B}$ function,
which is of order $\as^3(p_{\text{\scalefont{0.77}T}})$, adds the
relevant (singular) contributions necessary to reach NNLO accuracy
\cite{Monni:2019whf}, with regular contributions in $\pt$ being
subleading at this order.

The function $D$, which includes the relevant singular terms in
$\pt{}$, is derived from a suitable modification of the $\pt{}$
resummation formula, as explained in section 4 of
\citere{Monni:2019whf},
\begin{align}
\label{eq:resum}
{\rm d}\sigma_{\scriptscriptstyle\rm F}^{\rm res}=\frac{{\rm d}}{{\rm
    d}\pt}\left\{e^{-S}\mathcal{L}\right\}=e^{-S}\underbrace{\left\{-S^\prime\mathcal{L}+\mathcal{L}^\prime\right\}}_{\equiv
  D}\,,
\end{align}
where $\mathcal{L}$ denotes the luminosity factor up to NNLO,
including the convolution of the collinear coefficient functions with
the parton distribution functions (PDFs) and the squared hard-virtual
matrix elements for \F{} production.  Note that here we do not
truncate \eqn{eq:minnlo} at $\as^3$ by evaluating
$\left(D-D^{(1)}-D^{(2)}\right)=D^{(3)}+\mathcal{O}(\as^4)$ as it was
done in the original formulation of \minnlo{} in
\citere{Monni:2019whf}.  Instead, we preserve the total derivative in
\eqn{eq:resum} by keeping the respective terms beyond
$\mathcal{O}(\as^3)$, which was proposed in \citere{Monni:2020nks} as
a way to achieve a better agreement with fixed-order NNLO results by
accounting for subleading logarithmic contributions.  Finally, the
factor $F^{\rm corr}$ in \eqn{eq:bbar} ensures that
$\left(D-D^{(1)}-D^{(2)}\right)$, which has Born-like kinematics, is
appropriately spread in the \FJ{} phase space when generating \FJ{}
\POWHEG{} events \cite{Monni:2019whf}.

We have implemented our \minnlo{} generator for \wz{} production in
the \POWHEGBOXRES{} framework \cite{Jezo:2015aia}. Since no generator
for \wz{}+jet production was available, we have first implemented this
process in the \POWHEGBOXRES{} framework.  In a second step, we have
upgraded this implementation by means of the \minnlo{} method to
achieve NNLO QCD accuracy for \wz{} production, using the general
\minnlo{} implementation for colour-singlet production developed in
\citere{Lombardi:2020wju}.  As far as the physical amplitudes are
concerned, our calculation relies on
\OpenLoops{}~\cite{Cascioli:2011va,Buccioni:2017yxi,Buccioni:2019sur}
for all tree-level and one-loop amplitudes via the interface developed
in~\citere{Jezo:2016ujg}, while for the two-loop amplitudes
\textsc{VVamp}\,\cite{Gehrmann:2015ora,hepforge:VVamp} is used through
the interface to \Matrix{}\,\cite{Grazzini:2017mhc} developed in
\citere{Lombardi:2020wju}.  As for $ZZ$ production in
\citere{Buonocore:2021fnj}, we exploit the possibility of reweighting
events at the generation level (stage 4) to include the two-loop
contribution, since the evaluation of the two-loop helicity amplitudes
for massive diboson processes is known to be computationally very
demanding.  This procedure substantially reduces the computing time of
the process, since the two-loop contribution is just included at
generation level and evaluated once per (accepted) event. This feature
of the code can be controlled by appropriate settings of the
\texttt{run\char`_mode} flag, as described in detail in
\citere{Buonocore:2021fnj}.

Our calculation involves the evaluation of several convolutions with
the PDFs, for which we employ \noun{hoppet}~\cite{Salam:2008qg}.  More
precisely, the PDFs are read through the \textsc{lhapdf}
interface~\cite{Buckley:2014ana} and evolved internally by
\textsc{hoppet}~\cite{Salam:2008qg} as described in
\citere{Monni:2019whf}.  The evaluation of the polylogarithms entering
the collinear coefficient functions is done through the \noun{hplog}
package~\cite{Gehrmann:2001pz}.

In the following, we briefly summarize the most relevant technical
settings that we have used to generate NNLO$_{\rm QCD}$+PS events for
\wz{} production: For more detailed information on those settings we
refer the reader to \citeres{Monni:2020nks,Mazzitelli:2021mmm}.  At
large $\pt$, spurious contributions from higher-order logarithmic
corrections are avoided by using a modified logarithm introduced in
eq.\,(4.15) of \citere{Mazzitelli:2021mmm}.  For the renormalization
and factorization scales we employ the typical \minnlo{} scale setting
at small $\pt$, which is defined in eq.\,(14) of
\citere{Monni:2020nks}, while in the NLO \wz{}+jet cross section the
scale setting is changed to the one in eq.\,(19) of
\citere{Monni:2020nks} at large $\pt$ by activating the option {\tt
  largeptscales\,1}. Note that we choose $\Q =0$\,GeV in those
equations and the Landau singularity is regulated by freezing the
strong coupling and the PDFs for scales below $0.8$\,GeV.  Finally,
the option \texttt{doublefsr\,1} of the \POWHEGBOX{} is turned on, see
\citere{Nason:2013uba} for details.  As far as the parton shower is
concerned we use \PYTHIA{8}~\cite{Sjostrand:2014zea} with standard
settings, which implies a global recoil scheme for initial state
radiation (\texttt{SpaceShower:dipoleRecoil\,0}).

\subsection{NLO$_{\rm EW}$+PS implementation}
\label{sec:nloewps}
For the computation of NLO EW corrections we have implemented a
separate generator within \POWHEGBOXRES{} for \wz{} production, which
is capable of computing NLO$_{\rm EW}$+PS, NLO$_{\rm QCD}$+PS, and
combined NLO$_{\rm QCD+EW}$+PS corrections, all consistently matched
to QCD and QED parton showers for all massive diboson processes. This
implementation makes use of the \POWHEGBOXRES{}~\cite{Jezo:2015aia}
framework which allows for resonance-aware NLO subtraction and
matching. Within this framework we construct the relevant resonance
information for \wz{} production, and use the standard resonance
projectors of~\citere{Jezo:2015aia}. Also here we employ tree-level
and one-loop amplitudes from
\OpenLoops{}~\cite{Cascioli:2011va,Buccioni:2017yxi,Buccioni:2019sur}. This
generator is essentially equivalent to the ones developed
in~\citere{Chiesa:2020ttl}.

\subsection{Veto procedure for QCD and QED radiation}
\label{sec:veto}
In order to match both NNLO$_{\rm QCD}$ and NLO$_{\rm EW}$ predictions
consistently with QCD and QED parton showers in \PYTHIA{8} we use a
veto procedure similar to the one described
in~\citere{Granata:2017iod} (see Appendix D). In particular, we let
both the QCD and QED showers radiate in the entire phase space
(restricted only by the kinematical bound) by setting
\begin{align}
&\text{\tt pythia\char`.readString("SpaceShower:pTmaxMatch = 2")}\nonumber\,,	\\
&\text{\tt pythia\char`.readString("TimeShower:pTmaxMatch = 2")}\nonumber\,,
\end{align}
which sets the shower starting scale equal to the partonic energy
$\sqrt{s}$ of the event. For each showered event we perform an a
posteriori check of the shower history and we veto events that are not
consistent with the emissions generated by \POWHEG{} at Les Houches
Event (LHE) level.

More precisely, when computing NNLO$_{\rm QCD}$+PS predictions, we
need to restrict the QCD emissions generated by the shower, as
commonly done in the \POWHEG{} framework, while QED radiations remain
unconstrained, so that the entire kinematically allowed phase space is
covered. In order to do so, once an event is showered we scan all the
QCD emissions generated by \PYTHIA{8}, store the hardest transverse
momentum $p_{\rm T}^{\rm max}$ and compare it to the hardness of the
QCD emission generated by \POWHEG{} (commonly referred to as {\tt
  scalup}), whose value is read from the event file. If $p_{\rm  T}^{\rm max}$
is greater than {\tt scalup}, we reject the event and
try to shower it again until the above requirement is fulfilled. After
1000 unsuccessful attempts the event is rejected.

For the generation of NLO$_{\rm EW}$+PS predictions the QED shower
must be restricted, while QCD radiation is unconstrained. QED
emissions can be generated both in the production of the two vector
bosons and in their resonance decays. Therefore, the shower has to be
vetoed both in the production and in the resonance decays of the
vector bosons using different veto scales. To do so, we generate
events according to the multiple-radiation scheme ({\tt allrad 1}),
first introduced in~\citere{Jezo:2016ujg}, which allows us to
distinguish between the generation of radiation from each QED-singular
region of the process at hand. In particular, up to one photon
emission can be generated in the production stage through
initial-state radiation (ISR) and up to one photon can be radiated
from each decaying resonance as final-state radiation (FSR). After the
event is showered, we scan the list of QED emissions generated by
\PYTHIA{8} and we store the transverse momenta of the hardest
emissions in the three regions. We then construct our veto scales: For
the production stage we store the transverse momentum of the photon
generated by \POWHEG{} as ISR, while for the two resonances we
calculate the transverse momentum of the photon generated at LHE level
with respect to the lepton emitter in the centre-of-mass frame of the
mother resonance, thus, obtaining two different scales for the two
vector bosons. If no photon is produced by \POWHEG{} in a certain
region, the corresponding veto scale is set equal to an infra-red
cutoff ($10^{-3}\,$ GeV). For each region we check whether the shower
contains QED emissions harder than the constructed veto scale in that
region. If that is the case, we veto the event and we try to shower it
again. After 1000 attempts, the event is rejected.

\subsection{Combination of NNLO QCD and NLO EW corrections}
\label{sec:combination}

In this paper, NNLO$_{\rm QCD}$+PS and NLO$_{\rm EW}$+PS events are
generated and showered with \PYTHIA{8} separately, and their
combination is performed a posteriori at the level of differential
distributions.

There is a number of different ways how these combinations of
higher-order QCD and EW predictions can be defined. First, the QCD and
EW perturbative corrections can either be added or multiplied. In the
high-energy regime, i.e.\ in situations where EW effects are dominated
by EW Sudakov logarithms~\cite{Denner:2000jv,Accomando:2004de}, and
when the dominant QCD effects arise at scales well below the hard
scale, a multiplicative combination should be seen as superior, as
such QCD effects factorise with respect to the underlying hard diboson
process. However, this assumption is violated in the
phenomenologically relevant situation where the process is dominated
by underlying hard vector-boson plus jet topologies with an additional
soft vector boson. These configurations are forbidden at LO and lead
to $\mathcal{O}(1)$ NLO QCD corrections, known as \textit{giant
  K-factors}~\cite{Rubin:2010xp,Baglio:2013toa}. In this regime a
multiplicative combination of QCD and EW effects overestimates the
impact of the EW corrections as those EW corrections determined for
the hard diboson process are applied to the hard vector-boson plus jet
topologies. In turn, an additive combination will largely
underestimate the EW corrections, as in this case no EW corrections
are considered for the dominating vector-boson plus jet topologies. As
discussed in \citere{Grazzini:2019jkl} the average of a multiplicative
and additive combination can be considered as a pragmatic estimate in
such situations. However, as also pointed out in
\citere{Grazzini:2019jkl}, when one is interested in the hard diboson
process, in general it is advisable to avoid such topologies through
appropriately defined vetos of hard QCD radiation, ideally defined
dynamically in phase space.

Second, in any combination a double-counting of both QCD and QED
radiation has to be avoided.  In this regard one may choose whether
QCD and/or QED emissions in the parton shower are accounted for in
both the NNLO$_{\rm QCD}$+PS and the NLO$_{\rm EW}$+PS calculation, or
whether either the QED shower is turned off in the NNLO$_{\rm QCD}$+PS
calculation, or the QCD shower is turned off in the NLO$_{\rm EW}$+PS
calculation. Any of these choices is consistent as long as the desired
formal accuracy is reached without double counting, namely NNLO$_{\rm
  QCD}$ and NLO$_{\rm EW}$ in the perturbative expansion and
leading-logarithmic accuracy in both QCD and QED shower emissions.

In order to distinguish the relevant combination schemes we introduce
the following notation: NNLO$_{\rm QCD}^{\rm (QCD,QED)_{\rm PS}}$
refers to NNLO accuracy in QCD perturbation theory matched to a parton
shower that includes both QCD and QED emissions, while NNLO$_{\rm
  QCD}^{\rm (QCD)_{\rm PS}}$ corresponds to the same perturbative
accuracy but with QED shower turned off. Similarly, we introduce
NLO$_{\rm EW}^{\rm (QCD,QED)_{\rm PS}}$ and NLO$_{\rm EW}^{\rm
  (QED)_{\rm PS}}$ for the NLO cross section in the EW expansion with
and without QCD shower, respectively, as well as the corresponding
symbols at LO with both or either one of the two showers turned on,
i.e.\ LO$^{\rm (QCD,QED)_{\rm PS}}$, LO$^{\rm (QCD)_{\rm PS}}$, and
LO$^{\rm (QED)_{\rm PS}}$.  Furthermore, we introduce a generic term
$\delta$N(N)LO$_{\rm X}^{\rm (Y,Z)_{\rm PS}}$ for the coefficient of
the $\rm X=\{\rm QCD,EW\}$ higher-order correction defined as
\begin{equation}
  \delta{\rm N(N)LO}_{\rm X}^{\rm (Y,Z)_{\rm PS}} = {\rm N(N)LO}_{\rm X}^{\rm (Y,Z)_{\rm PS}} - {\rm LO}_{\rm X}^{\rm (Y,Z)_{\rm PS}}\,,
  \end{equation}
and a multiplicative correction factor $\text{K-N(N)LO}_{\rm X}^{\rm (Y,Z)_{\rm PS}}$, which reads
\begin{equation}
\text{K-N(N)LO}_{\rm X}^{\rm (Y,Z)_{\rm PS}} =  {\rm N(N)LO}_{\rm X}^{\rm (Y,Z)_{\rm PS}} /{\rm LO}_{\rm X}^{\rm (Y,Z)_{\rm PS}}\,.
\end{equation}
We also define a corresponding NLO EW correction factor $\text{K-NLO}_{\rm EW}^{\rm (f.o.)}$ obtained at fixed-order with \Matrix+\OpenLoops{}
\begin{equation}
\text{K-NLO}_{\rm EW}^{\rm (f.o.)} =  {\rm NLO}_{\rm EW}^{\rm (f.o.)} /{\rm LO}^{\rm (f.o)}\,.
\end{equation}

Adopting these notations, we introduce the following schemes to
combine NNLO QCD and NLO EW corrections matched to QCD and QED parton
showers:

\begin{align}
&\text{additive schemes:}\nonumber\\
&1.\;\addfull{} = {\rm NNLO}_{\rm QCD+EW}^{\rm (QCD, QED)_{\rm PS}}  \\
&2.\;\addqcdfull{}\\
&3.\;\addqedfull{}\\
&\text{multiplicative schemes:}\nonumber\\
&4.\;\multfull{}= {\rm NNLO}_{\rm QCD\times EW}^{\rm (QCD, QED)_{\rm PS}}\label{eq:best}\\
&\text{5. NNLO}_{\rm QCD}^{\rm (QCD,QED)_{\rm PS}} \times \text{K-NLO}_{\rm EW}^{\rm (QED)_{\rm PS}}\\
&\text{6. NLO}_{\rm EW}^{\rm (QCD,QED)_{\rm PS}} \times \text{K-NNLO}_{\rm QCD}^{\rm (QCD)_{\rm PS}}\\
&\text{7. } \QCDtEWfo, \label{NNLOPStEWfo}
\end{align}
where for the first and fourth combination we introduced the dedicated
short-hand notations \QCDpEW and \QCDtEW, respectively.  In the result
section we will consider these combination schemes and study which of
them are more appropriate than others based on their ability to
describe relevant distributions in the most accurate way.

\section{Phenomenological results}
\label{sec:results}

In the following we study phenomenological results for \wz{} at
NNLO$_{\rm QCD}$ and NLO$_{\rm EW}$ accuracy matched to parton
showers. For brevity and without loss of generality, we focus on the
process
\begin{align}
	p p  \to \mu^+\nu_\mu e^+e^- \,,
\end{align}
 but all qualitative conclusions apply also to the case of the charged
 conjugated process with an intermediate negatively charged $W$ boson
 as well as to same-flavour leptonic final states. Of course, when
 comparing to ATLAS data \cite{ATLAS:2019bsc} in
 \sct{sec:comparisondata} we consider both charges in the final state,
 i.e.\ \mbox{$p p \to \mu^\pm\nu_\mu e^+e^- $}, and account for all
 relevant leptonic final states.

\subsection{Input parameters and setup}
\label{sec:settings}

We present results for proton--proton collisions at the LHC with a
center-of-mass energy of 13\,TeV. The complex-mass
scheme~\cite{Denner:1999gp,Denner:2005fg} is employed throughout and
the electroweak (EW) inputs are chosen according to their PDG values
\cite{ParticleDataGroup:2020ssz}: $G_{\text{\scalefont{0.77}F}} =
1.16639 \times 10^{-5}$~GeV$^{-2}$, $\mw = 80.385$~GeV,
$\Gamma_{\text{\scalefont{0.77}W}} = 2.0854$~GeV,
$m_{\text{\scalefont{0.77}Z}} = 91.1876$~GeV,
$\Gamma_{\text{\scalefont{0.77}Z}} = 2.4952$~GeV, \mbox{$m_H =
  125$~GeV} and $\Gamma_H = 0.00407$~GeV.  The on-shell mass and width
of the top-quark are set to $m_t = 173.2$~GeV and $\Gamma_t
=1.347878$~GeV.  All other EW parameters are determined through the
$G_{\text{\scalefont{0.77}$\mu$}}$ scheme, in particular by computing
the EW coupling as~\cite{Buccioni:2019sur}
\begin{align}
	\alpha_{\text{\scalefont{0.77}$G_\mu$}} = \frac{\sqrt{2}}{\pi}\, G_{\text{\scalefont{0.77}F}} |(m_{\text{\scalefont{0.77}W}}^2 - i \Gamma_{\text{\scalefont{0.77}W}}
m_{\text{\scalefont{0.77}W}}) \sin^2 \theta_{\text{\scalefont{0.77}W}}|\,,
\end{align}
 and the EW mixing angle as 
 \begin{align}
\cos^2 \theta_{\text{\scalefont{0.77}W}} =
\frac{m_{\text{\scalefont{0.77}W}}^2 - i \Gamma_{\text{\scalefont{0.77}W}} \mw{}}{ m^2_{\text{\scalefont{0.77}Z}} - i \Gamma_{\text{\scalefont{0.77}Z}} \mz{}}\,.
\end{align}  
As PDFs we use the five-flavour NNPDF3.1~\cite{Ball:2017nwa} NNLO set
with $\as=0.118$, specifically the {\tt
  NNPDF31\char`_nnlo\char`_as\char`_0118\char`_luxqed}
set~\cite{Manohar:2016nzj,Manohar:2017eqh,Bertone:2017bme}.  The
central factorization and renormalization scales are set as discussed
in \sct{sec:minnlo} for the \minnlo{} \wz{} generator. For the
NLO$_{\rm EW}$+PS calculation and fixed-order NNLO$_{\rm QCD}$ results
we set them as
\begin{align}
	 \muF=\muR=\frac{1}{2}\left(\sqrt{m_{\text{\scalefont{0.77}$e^+e^-$}}^2+p^2_{\text{\scalefont{0.77}T,$e^+e^-$}}}+\sqrt{m_{\text{\scalefont{0.77}$\mu\nu_\mu$}}^2+p^2_{\text{\scalefont{0.77}T,$\mu\nu_\mu$}}}\right)\,,
\end{align}
where $m_{\text{\scalefont{0.77}$e^+e^-$}}$ and
$p_{\text{\scalefont{0.77}T,$e^+e^-$}}$
($m_{\text{\scalefont{0.77}$\mu\nu_\mu$}}$ and
$p_{\text{\scalefont{0.77}T,$\mu\nu_\mu$}}$) are the invariant mass
and the transverse momentum of the reconstructed $Z$ boson ($W$
boson), respectively.  Scale uncertainties in all cases are estimated
via seven-point scale variation, where $\muF$ and $\muR$ are varied
around their central values by a factor of two in either direction,
with the constraint $0.5 \leq \muR /\muF \leq 2$. When NNLO$_{\rm
  QCD}$+PS and NLO$_{\rm EW}$+PS results are combined, as described in
\sct{sec:combination}, scale variations are assumed to be
correlated. Hence, the construction of multiplicative EW correction
factors is almost scale independent, up to relative corrections of
order $\alpha$ due to $\muF$ variations.

For all predictions matched to a parton shower presented in this paper
we make use of \PYTHIA{8}~\cite{Sjostrand:2014zea} with the Monash
2013 tune~\cite{Skands:2014pea} (\texttt{py8tune 14} in the input
card).  To validate our calculation, we compare NNLO$_{\rm QCD}$+PS
and NLO$_{\rm EW}$+PS results at LHE level to NNLO$_{\rm QCD}$ and
NLO$_{\rm EW}$ fixed-order predictions obtained with
\Matrix~\cite{Grazzini:2017mhc,Grazzini:2019jkl}.

In order to prevent charged resonances to radiate photons and photons
to branch into lepton- or quark-pairs, we set the two flags
\texttt{TimeShower:QEDshowerByOther} and\linebreak
\texttt{TimeShower:QEDshowerByGamma} to \texttt{off}.  We define
dressed leptons by adding to the four-momentum of a charged lepton
$\ell$ the four-momenta of all photons within a distance \linebreak
$\Delta R_{\ell \gamma} = \sqrt{\Delta \phi_{\ell \gamma}^2 + \Delta
  \eta_{\ell \gamma}^2} < 0.1$, starting from the smallest $R_{\ell
  \gamma}$ among all lepton--photon combinations, and removing any
recombined photons from the list of final-state particles.

\renewcommand{\baselinestretch}{1.5}
\begin{table}[b]
\centering
  \begin{tabular}{l|cc}
     & \setupinclusive & \setupfiducial \\
    \toprule
              $Z$-mass window & $66$\,GeV$ < m_{\text{\scalefont{0.77}$\lz^+\lz^-$}} < 116$\,GeV & $| m_{\text{\scalefont{0.77}$\lz^+\lz^-$}} - \mz{} | < 10$\,GeV\\[0.1cm]
    lepton cuts &  & $\begin{array}{c}
    p_{T,e^\pm}>15\,{\rm GeV}, \quad \ptlw>20\,{\rm GeV}, \\[-0.15cm]      
 |\etal|<2.5,\quad \mtw> 30\,{\rm GeV}, \\[-0.15cm]
 \Delta R_{\text{\scalefont{0.77}$\lz^+\lz^-$}} >0.2, \quad \Delta R_{\text{\scalefont{0.77}$\lz^\pm\lw$}}>0.3  
 \end{array}$
 \end{tabular}
 \renewcommand{\baselinestretch}{1.0}
  \caption{Inclusive and fiducial cuts used to define the  phase space regions of the 
  \setupinclusive{} and the \setupfiducial{}~\cite{Aaboud:2016yus}. 
  Note that $e^\pm$ and $\mu$ refer to dressed leptons.}
   \label{tab:cuts}
\end{table}
 \renewcommand{\baselinestretch}{1.0}

When validating our calculation and studying different combinations of
QCD and EW corrections matched to QCD and QED parton showering, we
consider two different setups: an inclusive one, referred to as
\setupinclusive{}, with just a mass window for the $Z$ boson, which
avoids the photon-pole singularity, and one with a set of fiducial
cuts referred to as \setupfiducial{}. These setups are summarized in
\tab{tab:cuts}. The \setupfiducial{} corresponds to the one used in
the ATLAS analyses of \citere{Aaboud:2016yus} and
\citere{ATLAS:2019bsc}, and is the default setup implemented in the
\Matrix{} code for \wz{} production. Finally, when comparing to ATLAS
data in \sct{sec:comparisondata}, we exploit the corresponding {\sc
  Rivet} routines~\cite{Bierlich:2019rhm} provided on the {\tt
  HEPdata} webpage\footnote{{\tt
    https://www.hepdata.net/record/ins1720438}} to obtain the
distributions in the fiducial volume defined by the recent ATLAS
analysis of~\citere{ATLAS:2019bsc}.\footnote{In our analysis, used for
  the results presented in \sct{sec:comparisonmatching}, we have
  defined the transverse mass of the $W$ boson in tab.~\ref{tab:cuts}
  as $m_{\text{\scalefont{0.77}T,W}} =\sqrt{
    \left(E_{\text{\scalefont{0.77}T,{\lw}}}+E_{\text{\scalefont{0.77}T,$\nu_{\lw}$}}\right)^2
    - p_{\text{\scalefont{0.77}T,$\lw\nu_{\lw}$}}^2}$ with
  $E_{\text{\scalefont{0.77}T,$x$}}^2=m_{\text{\scalefont{0.77}$x$}}^2+p_{\text{\scalefont{0.77}T,$x$}}^2$. The
  Rivet analysis, employed for the results presented in
  \sct{sec:comparisondata}, uses the same definition but sets the mass
  of the dressed leptons to zero in this formula.}

  For simplicity we do not include effects due to hadronization or
  multi-particle interactions (MPI) anywhere, but in the comparison
  against the recent ATLAS results in
  section~\sct{sec:comparisondata}.

\subsection{Validation against fixed-order predictions}
\label{sec:validation}

We start by separately validating our NNLO$_{\rm QCD}$+PS and
NLO$_{\rm EW}$+PS calculations by comparing results at the LHE level
to fixed-order predictions from \Matrix{}+\OpenLoops{}.

\subsubsection{NNLO QCD}

\begin{figure}[htb]
\begin{center}\vspace{-0.2cm}
\begin{tabular}{cc}
\includegraphics[width=.31\textheight]{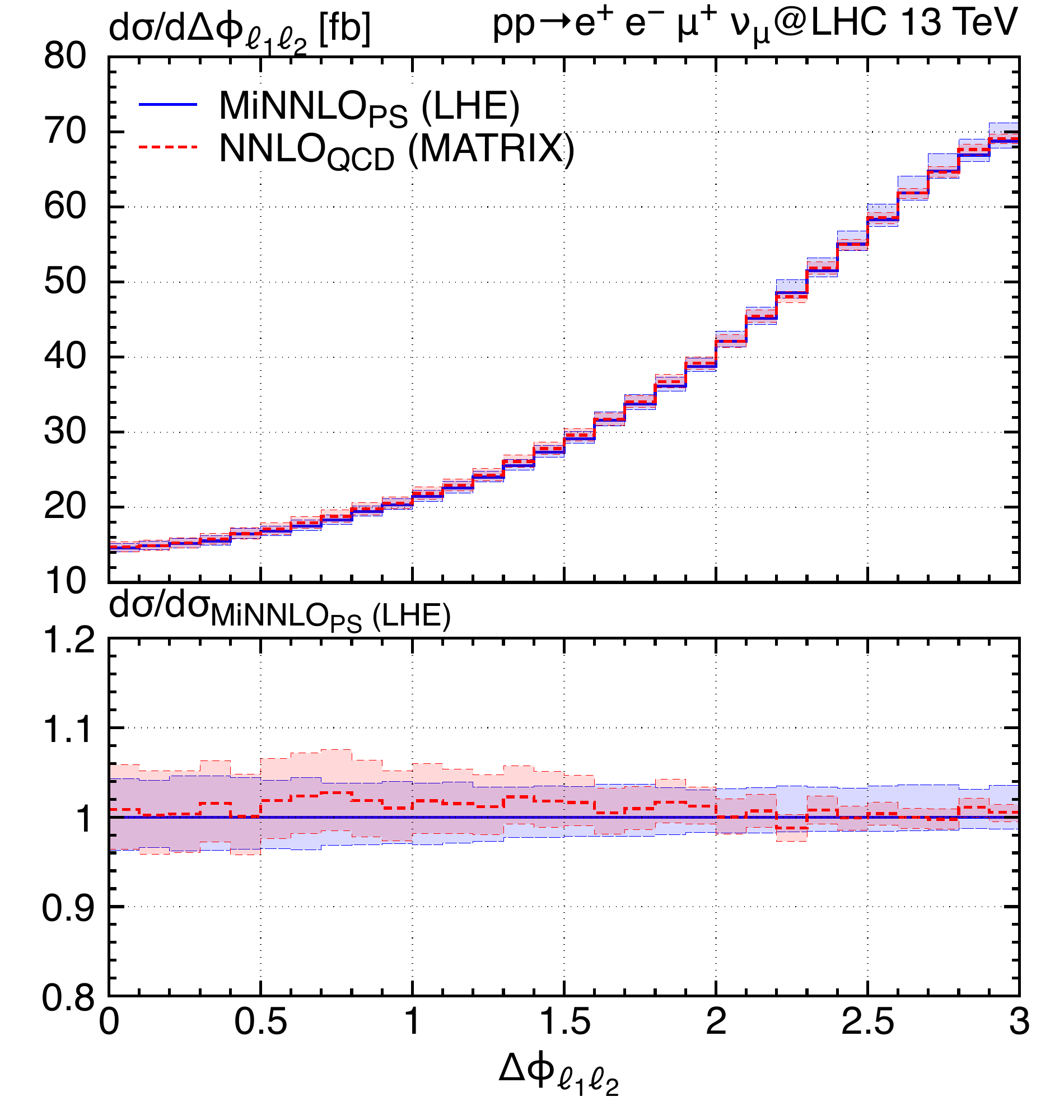} 
&
\includegraphics[width=.31\textheight]{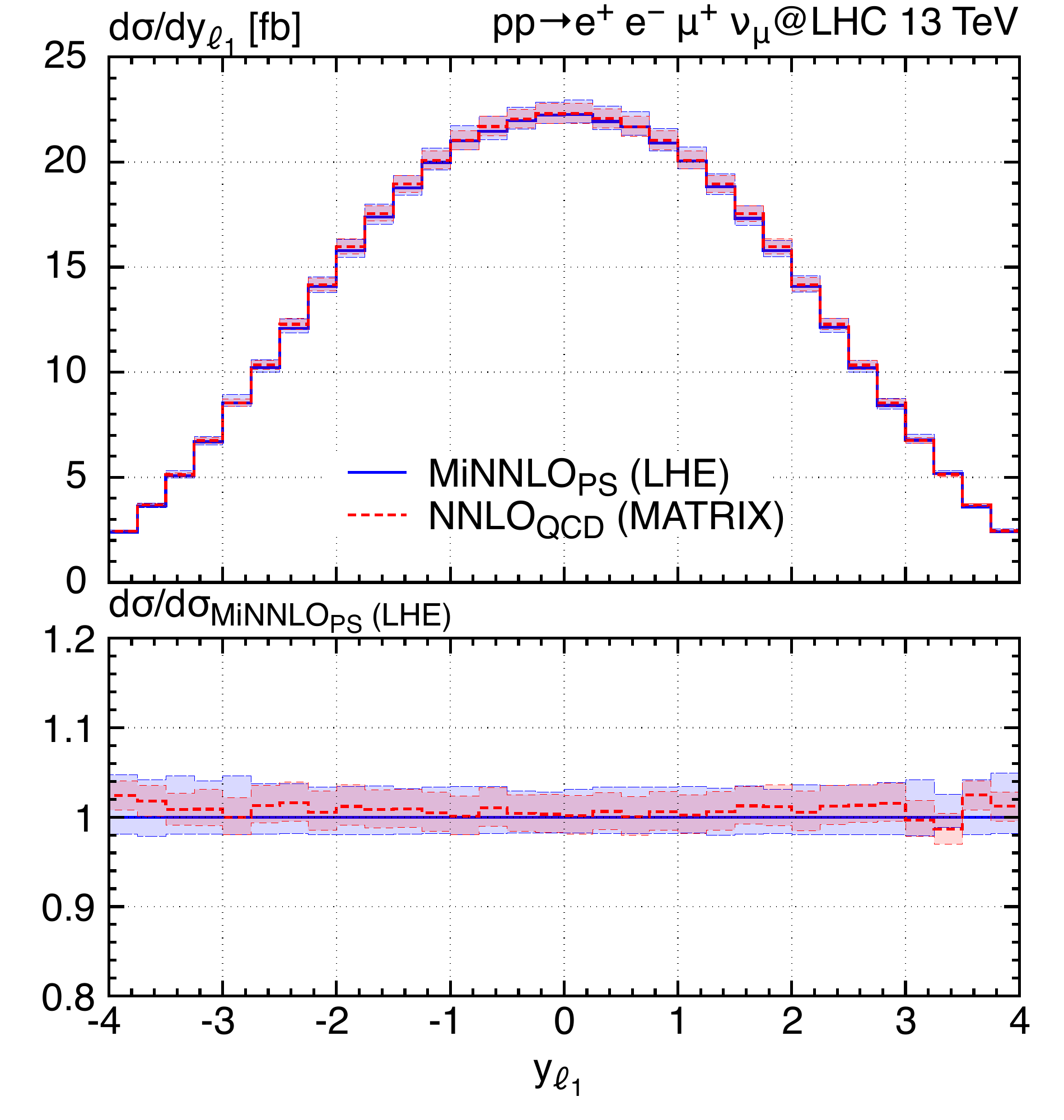}
\end{tabular}\vspace{-0.15cm}
\begin{tabular}{cc}
\includegraphics[width=.31\textheight]{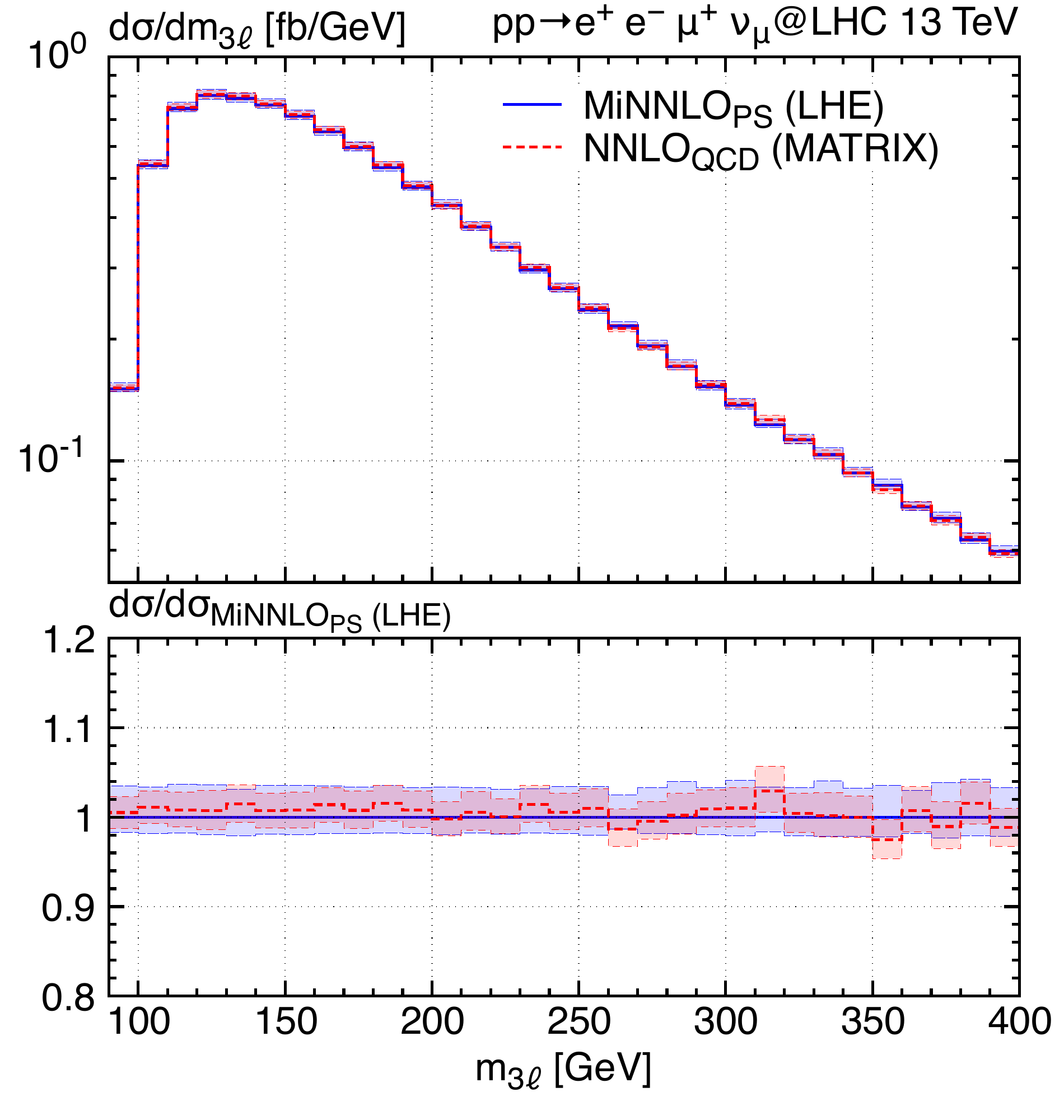}
&
\includegraphics[width=.31\textheight]{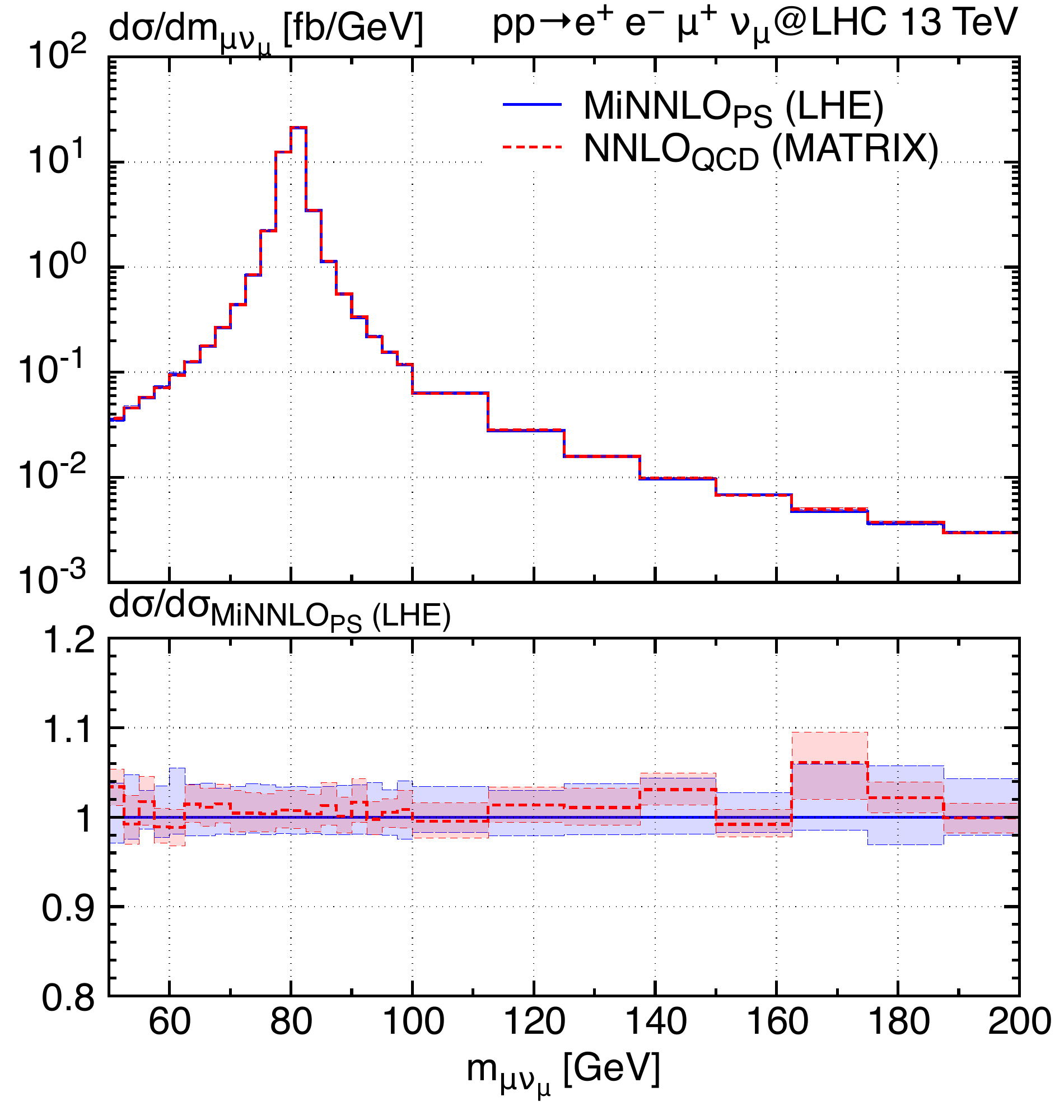}
\end{tabular}\vspace{-0.15cm}
\begin{tabular}{cc}
\includegraphics[width=.31\textheight]{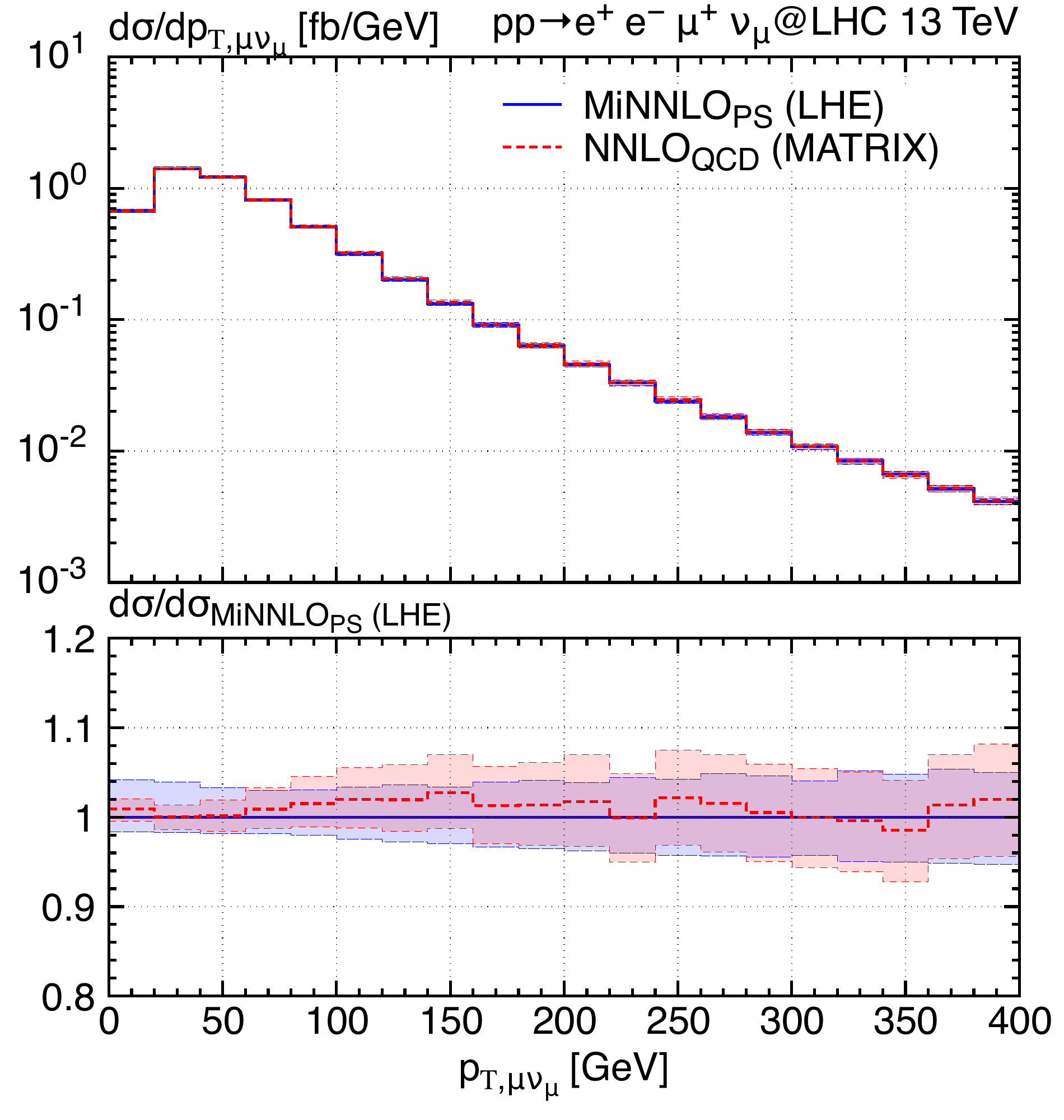}
&
\includegraphics[width=.31\textheight]{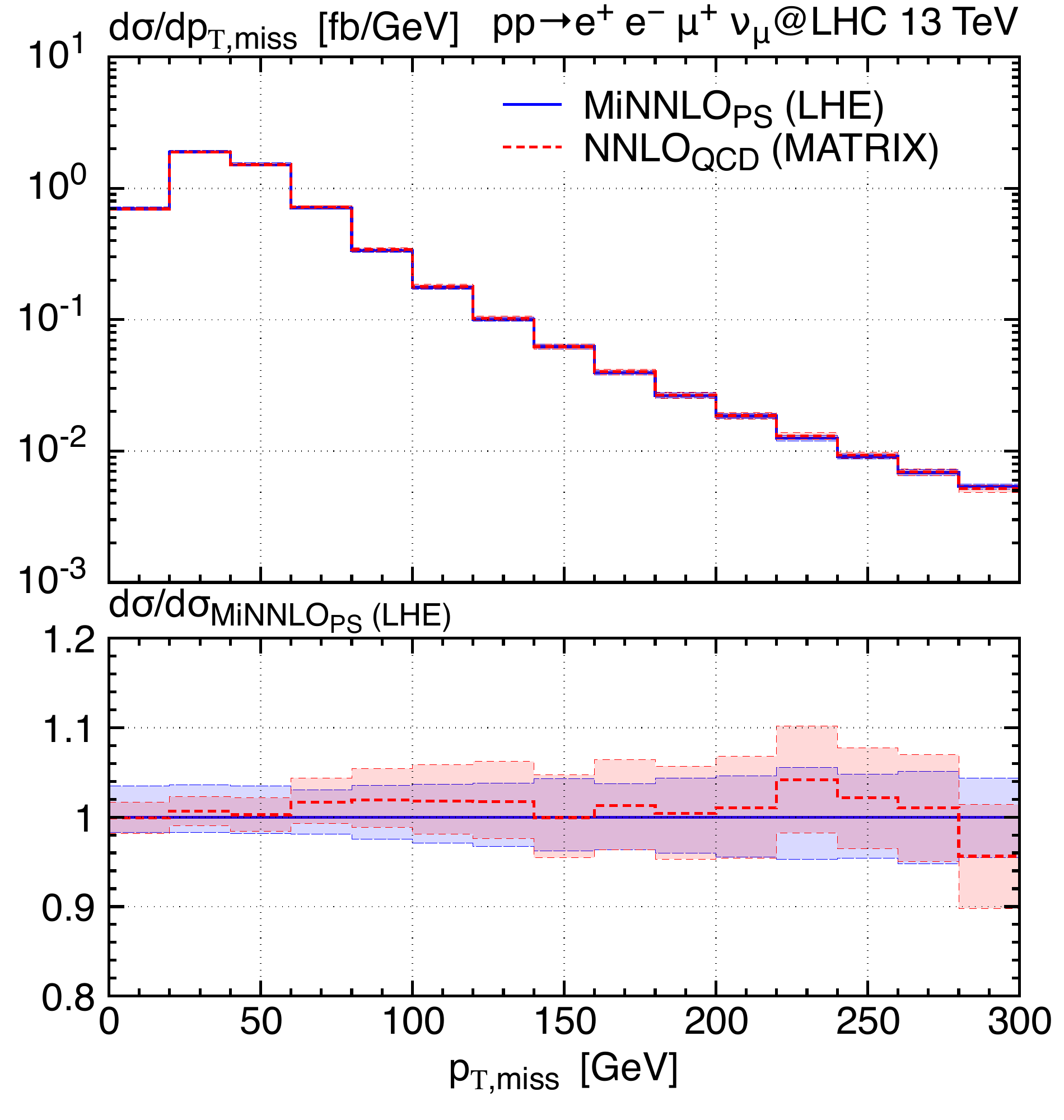} 
\end{tabular}
\caption{\label{fig:validationNNLO} Comparison of \minnlo{} (blue,
  solid) and fixed-order NNLO$_{\rm QCD}$ (red, dashed) predictions
  for $W^+Z$ production in the \setupinclusive{}.}
\end{center}
\end{figure}

In \fig{fig:validationNNLO} we show the comparison of NNLO$_{\rm
  QCD}$+PS predictions (blue, solid) at the LHE level obtained with
the \minnlo{} \wz{} generator to fixed-order NNLO$_{\rm QCD}$
predictions (red, dashed) for a selection of
distributions in the \setupinclusive{}. Specifically, we display the distributions in the
azimuthal difference between the leading and subleading charged lepton
($\dphill{}$), the rapidity of the leading charged lepton ($\ylone$),
the invariant mass of the three charged leptons ($\mlll$), the
invariant mass ($\mlnu$) and the
transverse momentum ($\ptlnu$) of the reconstructed $W$ boson, 
and the missing transverse momentum ($\ptmiss$). In all plots, the
main frame shows the distribution of the cross section in the
respective variable, while the lower panel shows the bin-by-bin ratio
of all curves to the \minnlo{} result.

As can be seen, for all distributions \minnlo{} predictions agree
nicely with the fixed-order NNLO$_{\rm QCD}$ reference result within
the perturbative uncertainties at NNLO. Also, the size of the
uncertainty bands are very similar between the two calculations. We
would like to stress that no one-to-one correspondence between the two
predictions is to be expected as they differ by higher-order terms,
both due to different choices in the treatment of terms beyond
accuracy and due to different scale settings.

We note that we have considered a large number of differential
observables and that we just show a representative selection of them
here. For all observables inclusive over QCD radiation, which are
genuinely NNLO$_{\rm QCD}$-accurate, \minnlo{} and fixed-order
NNLO$_{\rm QCD}$ results are in excellent agreement within the
respective scale uncertainties.  This comparison validates the
NNLO$_{\rm QCD}$ accuracy of our \minnlo{} generator, which we will
use in the following sections for phenomenological studies in
combination with higher-order EW effects.

\begin{figure}[t!]
\begin{center}\vspace{-0.2cm}
\begin{tabular}{cc}
\includegraphics[width=.31\textheight]{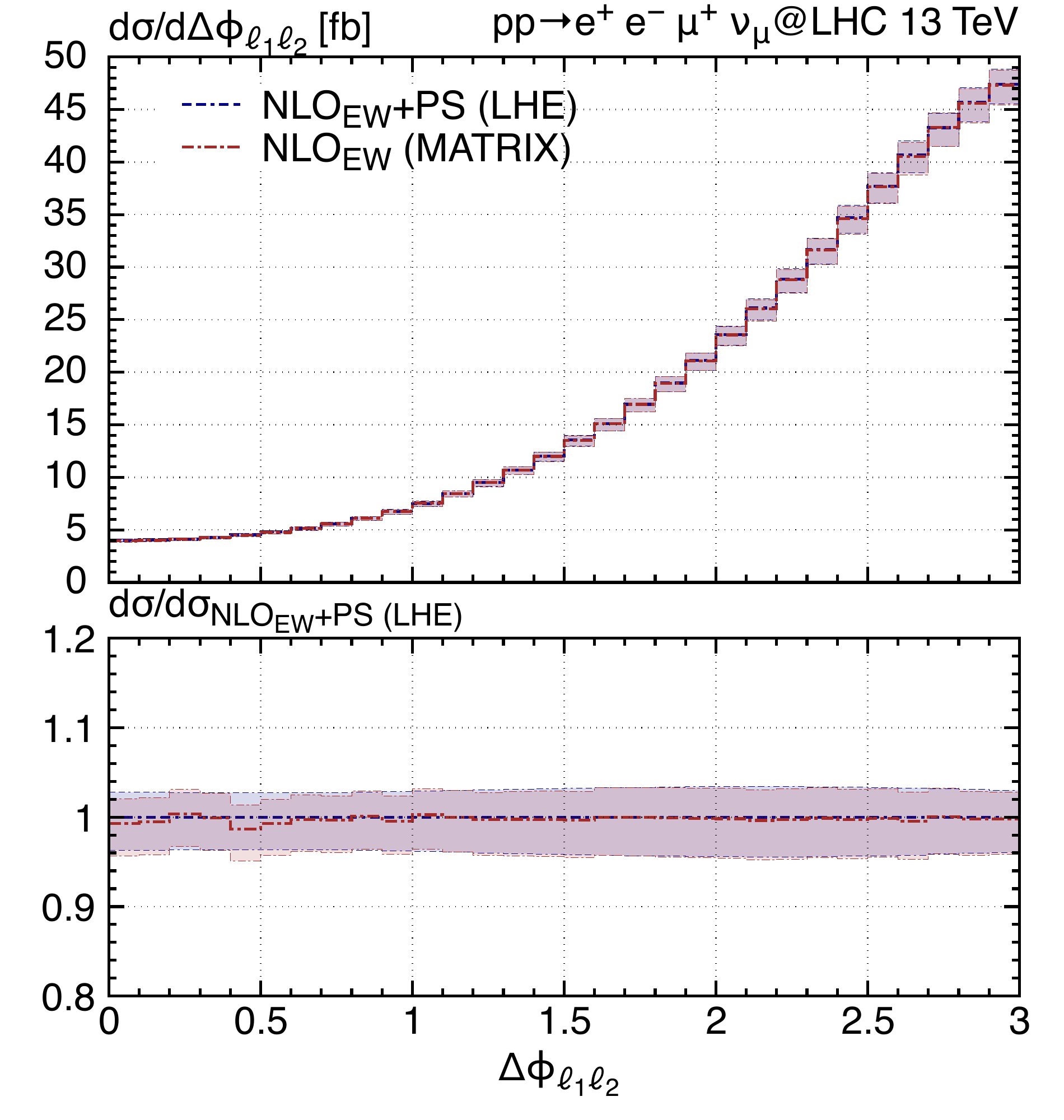}
&
\includegraphics[width=.31\textheight]{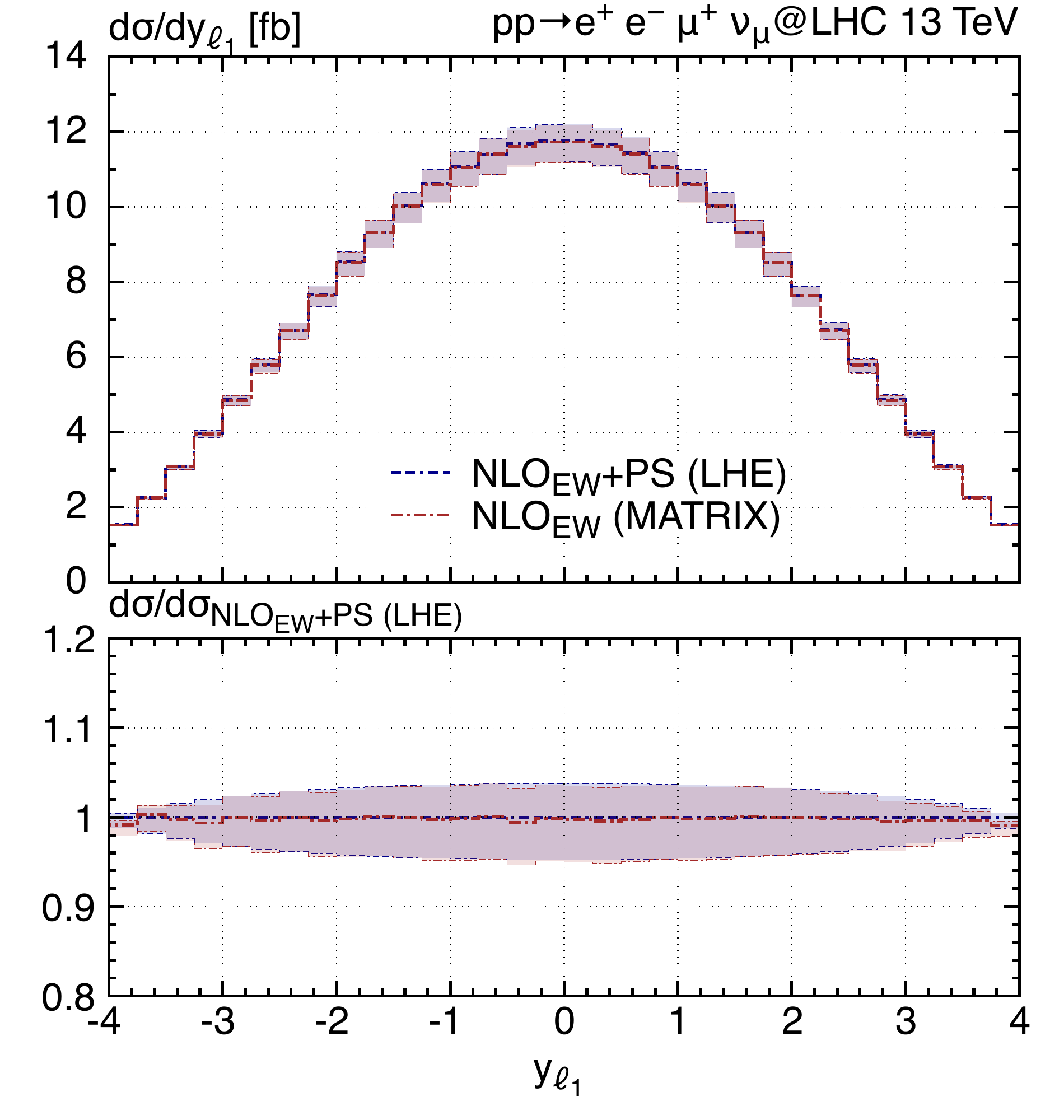}
\end{tabular}\vspace{-0.15cm}
\begin{tabular}{cc}
\includegraphics[width=.31\textheight]{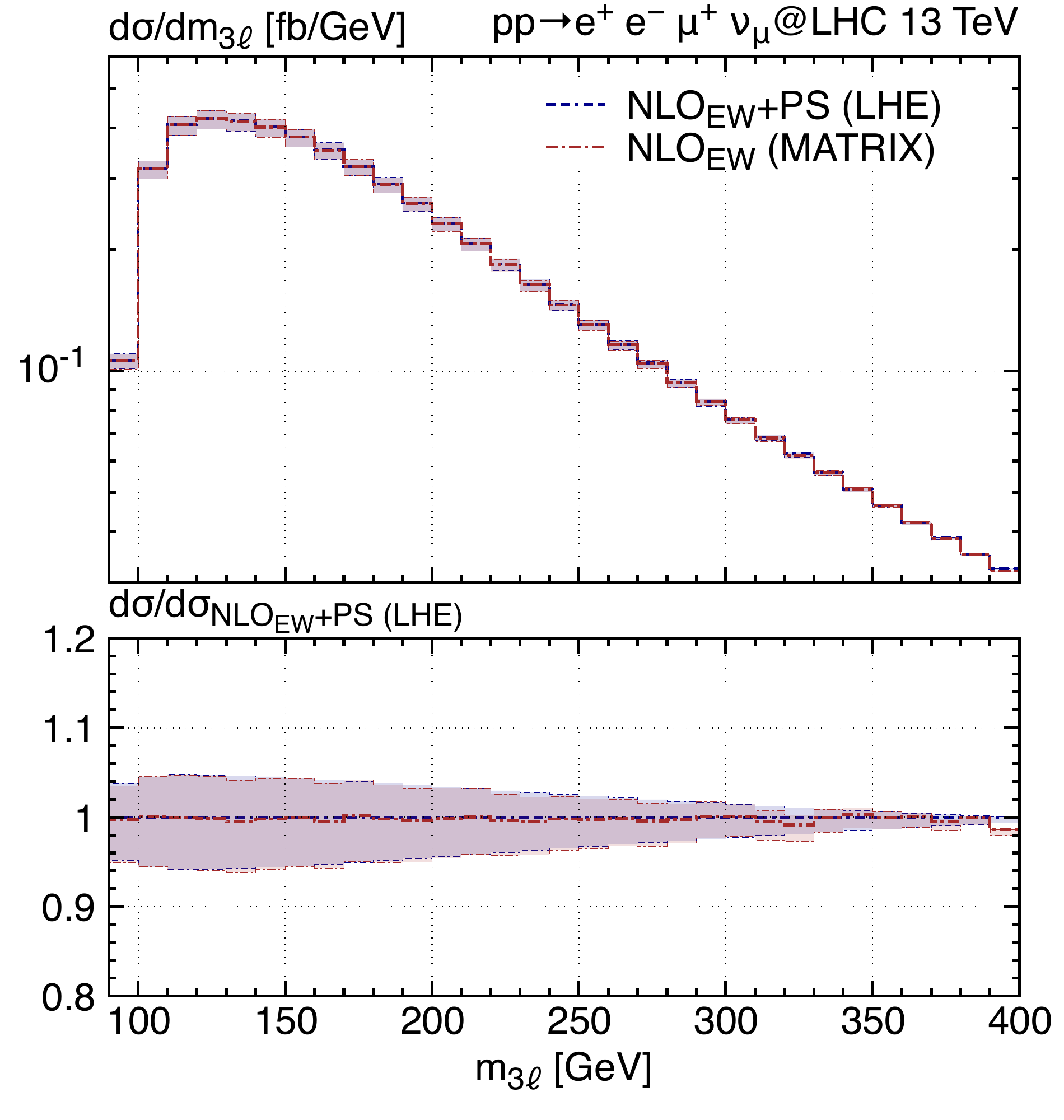} 
&
\includegraphics[width=.31\textheight]{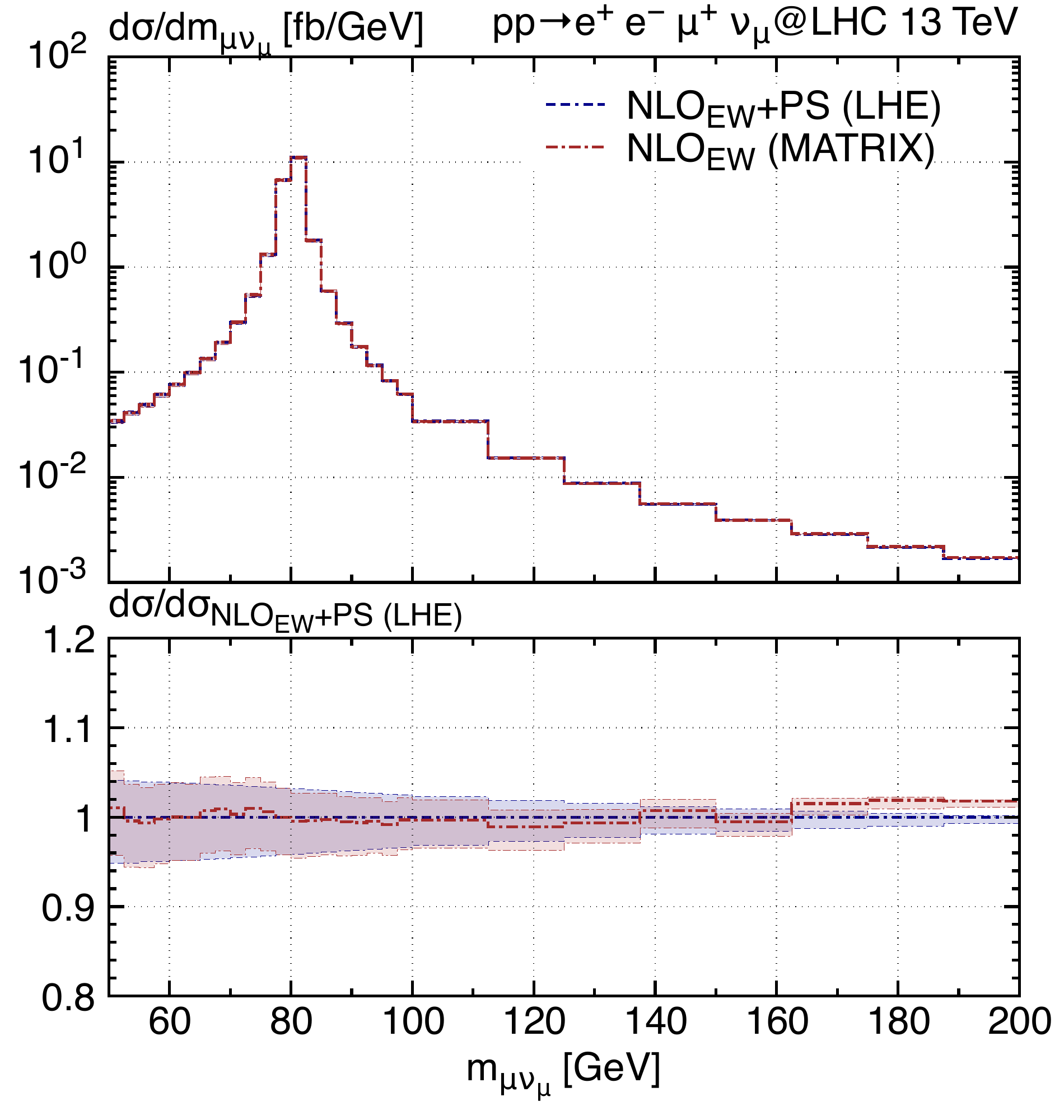}
\end{tabular}\vspace{-0.15cm}
\begin{tabular}{cc}
\includegraphics[width=.31\textheight]{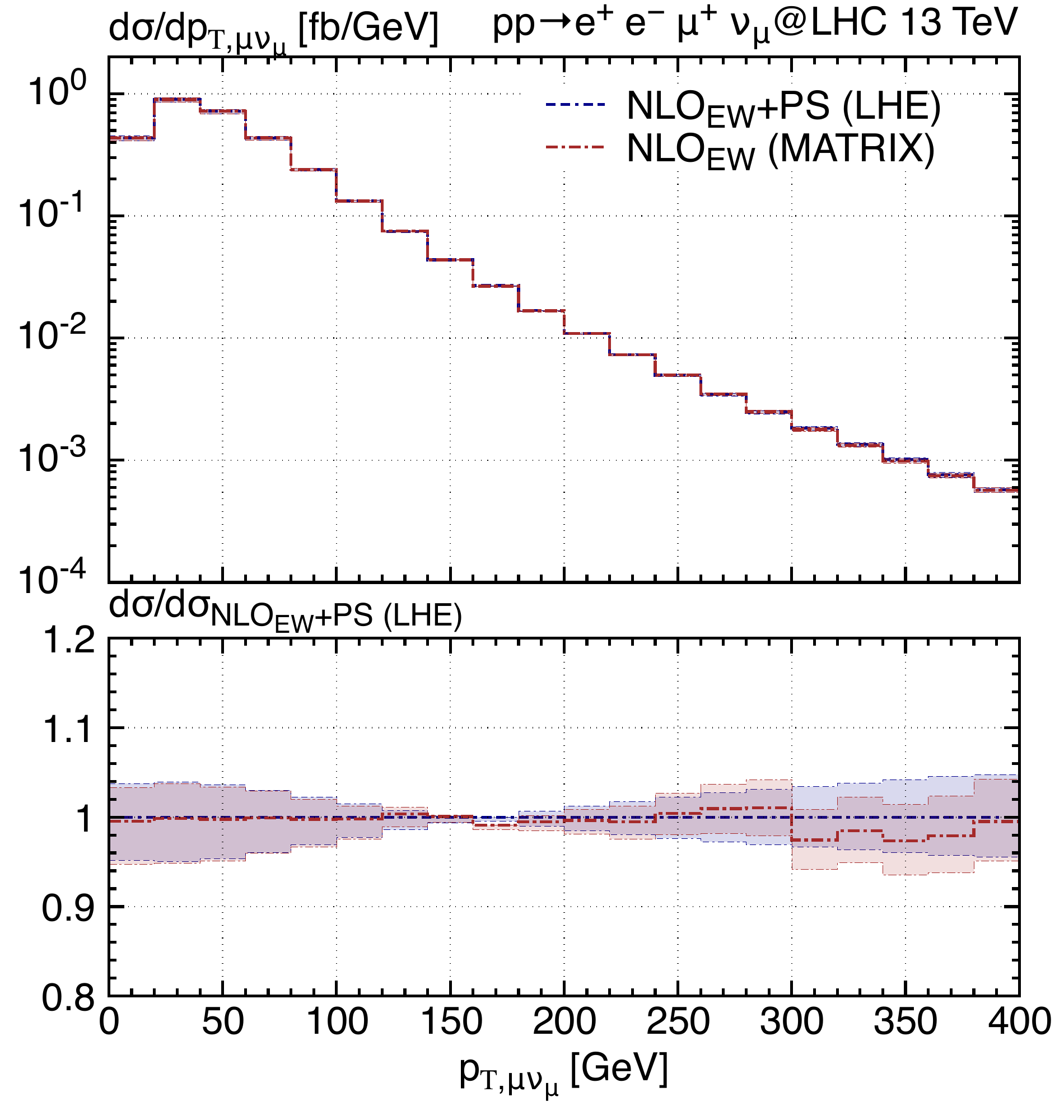} 
&
\includegraphics[width=.31\textheight]{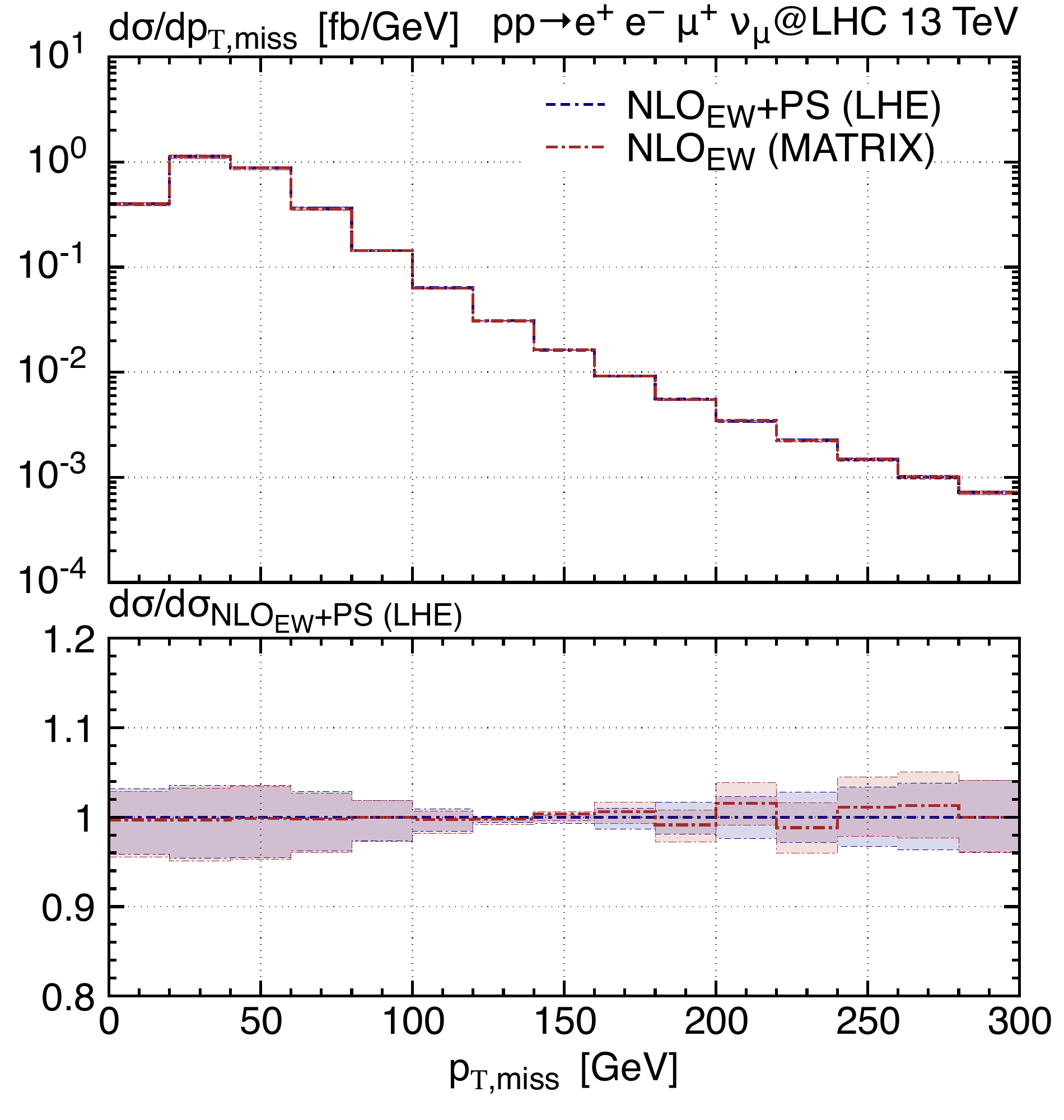}
\end{tabular}
\caption{\label{fig:validationEW} Comparison of NLO$_{\rm EW}$+PS
  (dark blue, double-dash dotted) and fixed-order NLO$_{\rm EW}$
  (brown, dash-dotted) predictions for $W^+Z$ production in the
  \setupinclusive{}.}
\end{center}
\end{figure}
\afterpage{\clearpage}

\subsubsection{NLO EW}

\Fig{fig:validationEW} validates our \POWHEG{} NLO$_{\rm EW}$
predictions at LHE level against fixed-order NLO$_{\rm EW}$ results
for the same set of observables discussed in the previous section in the \setupinclusive{}.
In this case, the two predictions do have a one-to-one correspondence in
the hard region and differ only by the contribution of the \POWHEG{}
Sudakov, which correctly distributes the first photon emission as
needed for the subsequent matching of the NLO$_{\rm EW}$ results at
LHE level to the parton shower. As a result, the \POWHEG{} NLO$_{\rm
  EW}$ (LHE) and fixed-order NLO$_{\rm EW}$ curves are essentially
identical up to numerical fluctuations in the tails of the
distributions. Moreover, also the size of the perturbative
uncertainties, which correspond essentially to LO in QCD, of either
prediction is practically identical.

Also in this case we have examined a large number of different
differential distributions, finding perfect agreement between the two
predictions for all NLO$_{\rm EW}$-accurate observables that are
inclusive over photon radiation. Therefore, we consider our NLO$_{\rm
  EW}$+PS generator fully validated as well, so that we can move on to
considering phenomenological results for the combination of NNLO$_{\rm
  QCD}$+PS and NLO$_{\rm EW}$+PS predictions in the next section.

\subsection{Parton-shower matched results}
\label{sec:comparisonmatching}

In this section we present results for \wz{} production at NNLO$_{\rm
  QCD}$ and NLO$_{\rm EW}$, both matched to QCD and/or QED parton
showers.  We consider different ways for the combination of NNLO$_{\rm
  QCD}$+PS and NLO$_{\rm EW}$(+PS) predictions, as introduced in
\sct{sec:combination}.  In order to distinguish between relevant EW
and QED effects, we compare these different combinations of NNLO$_{\rm
  QCD}$+PS and NLO$_{\rm EW}$+PS predictions to dedicated
approximations with lower formal accuracy, which include the pure
NNLO$_{\rm QCD}$+PS predictions with and without QED showering,
i.e.\ NNLO$_{\rm QCD}^{\rm (QCD, QED)_{\rm PS}}$ and NNLO$_{\rm
  QCD}^{\rm (QCD)_{\rm PS}}$, respectively, as well as a
multiplicative combination of NNLO$_{\rm QCD}$+PS results with a
fixed-order ${\rm EW}$ $K$-factor obtained through
\Matrix+\OpenLoops{}, as defined in \eqn{NNLOPStEWfo}.

Unless otherwise stated, all figures throughout this section are
organized as follows: The main frame shows \qcdfull{} (blue, dashed),
\QCDpEW{} (magenta, long-dashed) and \QCDtEW{} (green, solid)
predictions. In the first ratio panel these curves are normalized to
the \qcdfull{} prediction to visualize the EW effects.  In addition,
the \qcdqcd{} result (red, dash-dotted), i.e.\ NNLO QCD matched to a
QCD shower but without QED shower, is shown.  In the second ratio
inset the additive and multiplicative QCD--EW combinations are
normalized to the additive one to better display their
differences. Additionally, in this second ratio panel we show the
\QCDtEWfo combination (brown, dash-dotted), and, where explicitly
stated, further combinations as defined in
Section~\ref{sec:combination}. In all plots the uncertainty bands
correspond to seven-point scale variations, keeping scale variations
in the QCD and EW predictions correlated.  We stress that
multiplicative EW correction factors are essentially scale
independent, and that for all combination schemes the uncertainties
are dominated by NNLO QCD scale variations.

\begin{figure}[t]
\begin{center}\vspace{-0.2cm}
\begin{tabular}{cc}
\includegraphics[width=.31\textheight]{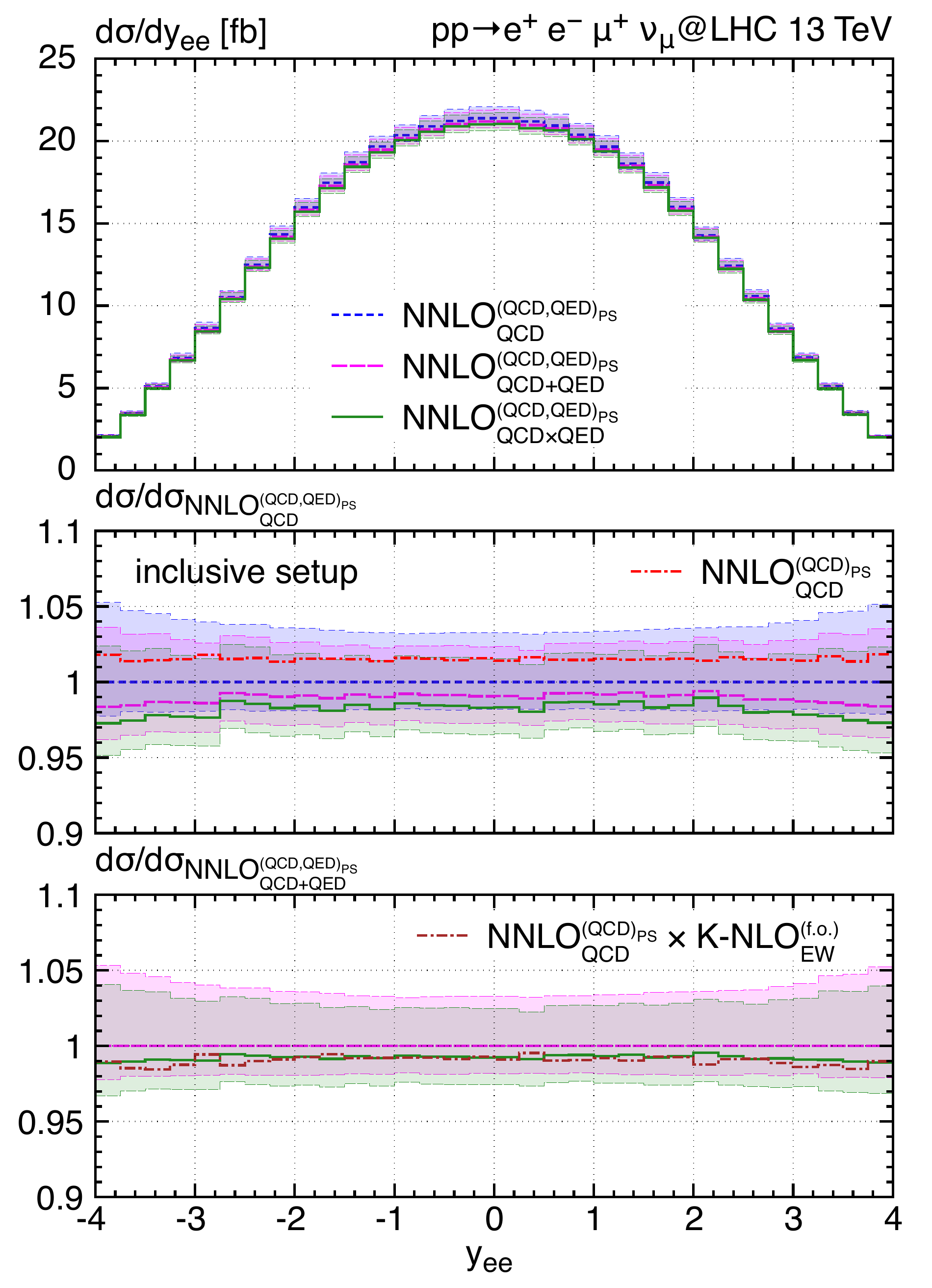}
&
\includegraphics[width=.31\textheight]{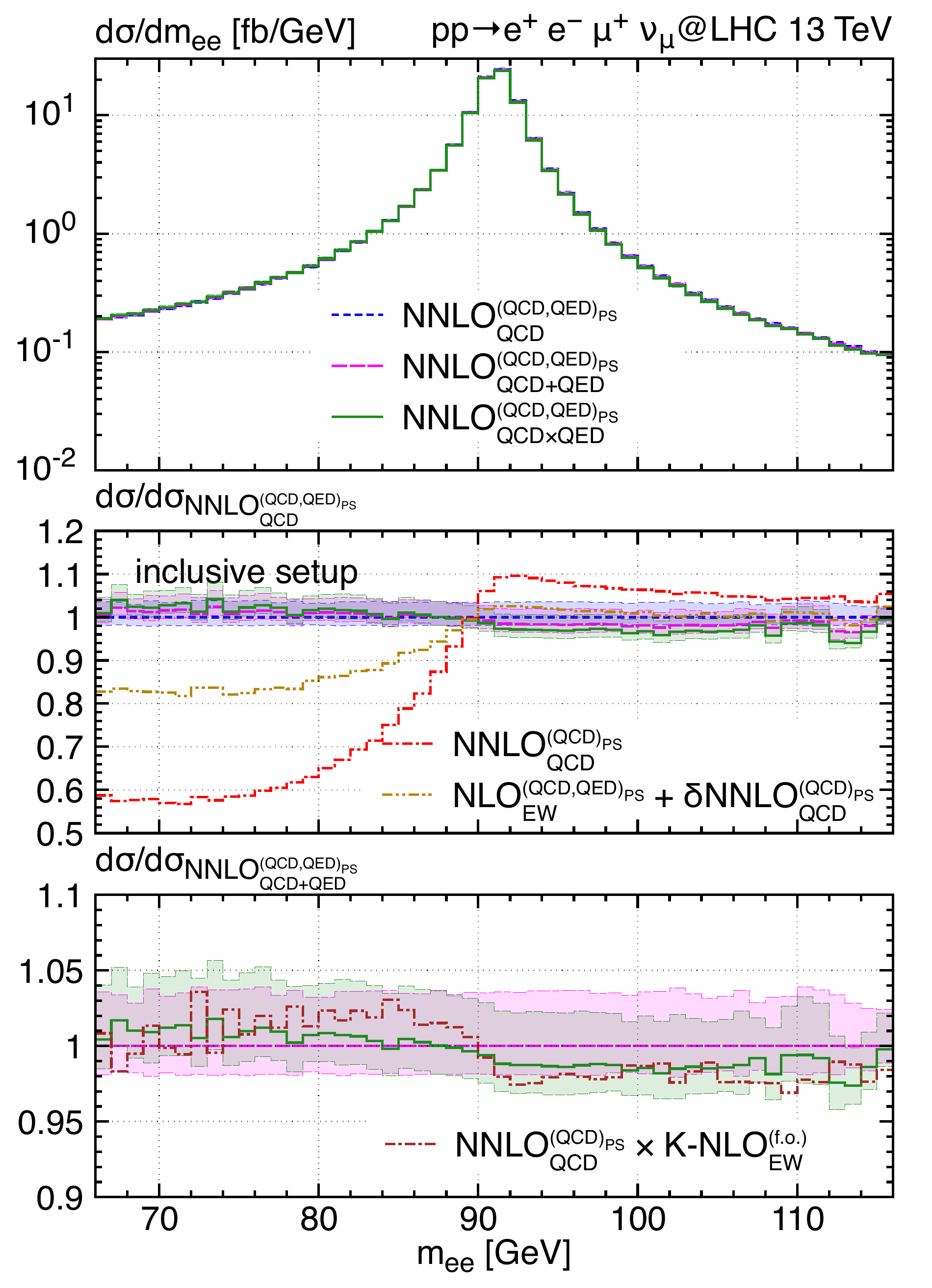}
\end{tabular}
\caption{\label{fig:combinations1} Differential distributions of the
  dilepton rapidity originating from the Z-boson (left) and of the
  corresponding dilepton invariant mass (right) in $W^+Z$ production in
  the \setupinclusive{} at NNLO$_{\rm QCD}$ combined with NLO$_{\rm
    EW}$ matched to parton showers for different combination
  schemes. See text for details.}
\end{center}
\end{figure}

We start our discussion of the numerical results in
\fig{fig:combinations1} with observables focusing on the reconstructed
$Z$-boson, namely the rapidity ($\yll$) and the invariant mass
($\mll$) of the lepton-pair associated with the $Z$ boson in the
\setupinclusive{}.  Looking at the $\yll$ distribution in
\fig{fig:combinations1}\,(left) we observe scale-uncertainty bands
with upper and lower edges at the level of $+ 3$--$5\%$ and $-
2$--$3\%$, respectively, in all shown predictions. EW corrections are
smaller than these QCD scale variations and show hardly any shape
effects, as expected from this observable that is inclusive with
respect to QED radiation. Indeed, comparing the \qcdqcd prediction
against the \qcdfull one indicates that pure QED effects are at the
level of \mbox{$-1$--$2\%$}, and an additional $-2$--$3\%$ of weak
origin is found when comparing further against the NLO EW-matched
\QCDpEW or \QCDtEW predictions, which in turn agree at the one percent
level. We also observe that the \QCDtEWfo prediction is practically
identical with the \QCDtEW one, which implies that multiple photon
emissions (beyond the first one) do not have a relevant impact here.

Looking at the $\mll{}$ distribution in
\fig{fig:combinations1}\,(right), the observations are different:
there are large effects from collinear QED radiation which shift
events from above the Breit–Wigner peak to below the peak. These
effects are entirely absent in the \qcdqcd prediction showing
deviations of up to 40\% compared to the \qcdfull{} prediction
including effects from the QED shower.  The observed shape of the
corrections due to these collinear QED effects is qualitatively very
similar to the well-known NLO EW corrections to neutral-current
Drell–Yan (plus jet) production for dressed
leptons~\cite{CarloniCalame:2007cd,Dittmaier:2009cr,Kallweit:2015dum}.
It is interesting to notice that the NNLO+PS QCD prediction with a QED
shower, i.e \qcdfull{}, provides an excellent approximation of this
distribution with respect to the full \QCDpEW and \QCDtEW
combinations, see central ratio panel in
\fig{fig:combinations1}\,(right). In the same panel, we also included
the central \addqedfull{} result (brown, dash-double-dotted) in order
to show that it does not provide a suitable prediction for the $\mll$
distribution, as it misses important mixed QED--QCD effects (although
beyond accuracy) originating from QED corrections relative to the NNLO
QCD contribution, and, thus, it can be discarded as a useful
combination from the list in \eqn{NNLOPStEWfo}.  Comparing the \QCDtEW
prediction with the \QCDtEWfo combination we observe agreement at the
$1$--$2\%$ level, see lower ratio panel in \fig{fig:combinations1}
(right). On the one hand, this further validates the employed
resonance-aware matching of the EW corrections with QED parton-shower
radiation within \POWHEGBOXRES{}, and, on the other hand, indicates
only a mild impact of multi-photon radiation beyond the first
emission, despite the sizable QED effects due to collinear photon
radiation.  Furthermore, given the very small observed differences
between the additive and multiplicative combination schemes at the
level of $1$--$2\%$, i.e.\ well below scale uncertainties, we consider
the difference between the two combination schemes a reliable estimate
of remaining mixed QCD--EW effects for the observable at hand.

\begin{figure}[t]
\begin{center}\vspace{-0.2cm}
\begin{tabular}{cc}
\includegraphics[width=.31\textheight]{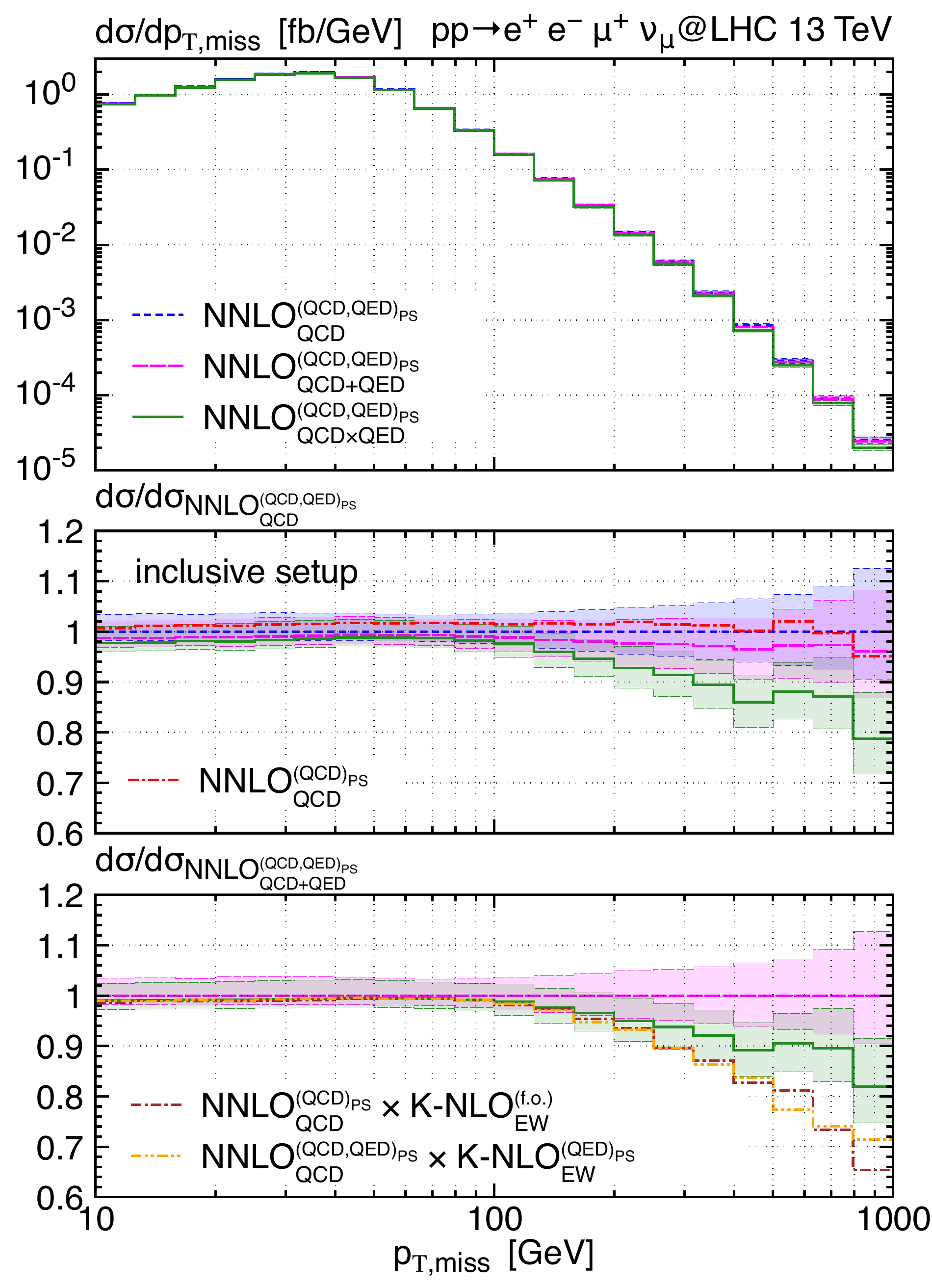}
&
\includegraphics[width=.31\textheight]{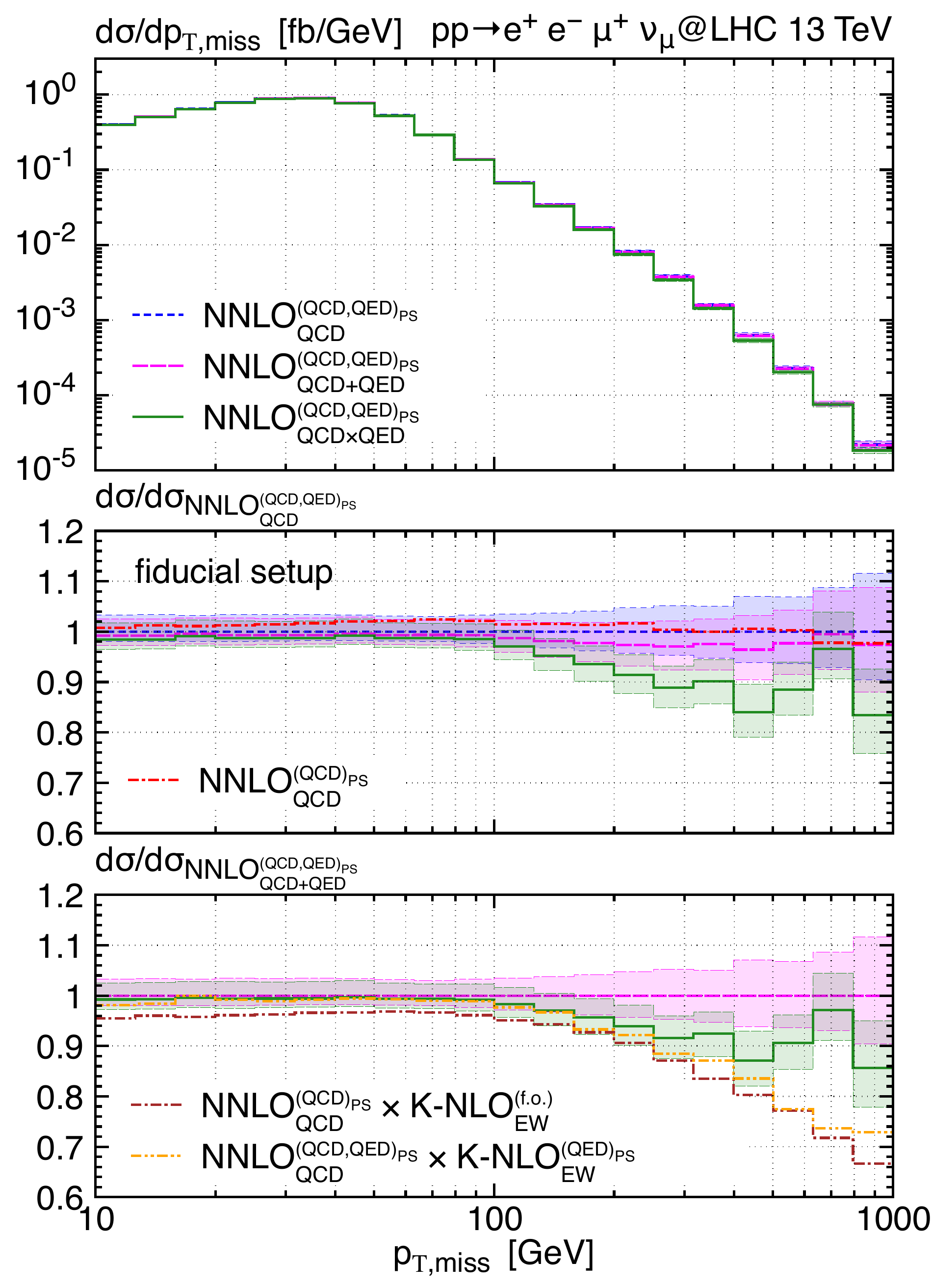} 
\end{tabular}
\caption{\label{fig:results_met} Differential distributions in the
  missing transverse-momentum $\ptmiss$ in $W^+Z$ production in the
  \setupinclusive{} (left) and \setupfiducial{} (right). See text for
  details.}
\end{center}
\end{figure}

\begin{figure}[th]
\begin{center}\vspace{-0.2cm}
\begin{tabular}{cc}
\includegraphics[width=.31\textheight]{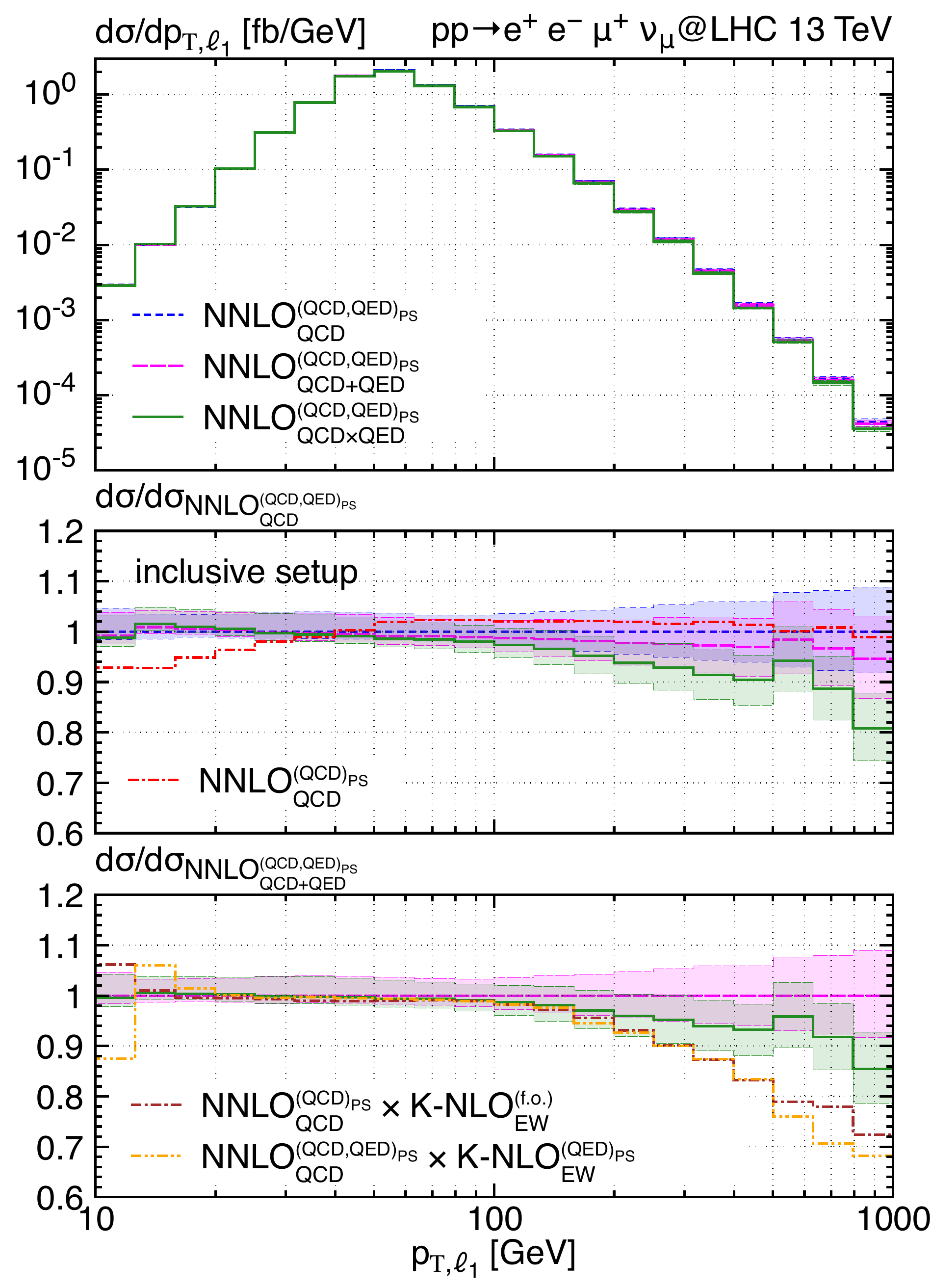}
&
\includegraphics[width=.31\textheight]{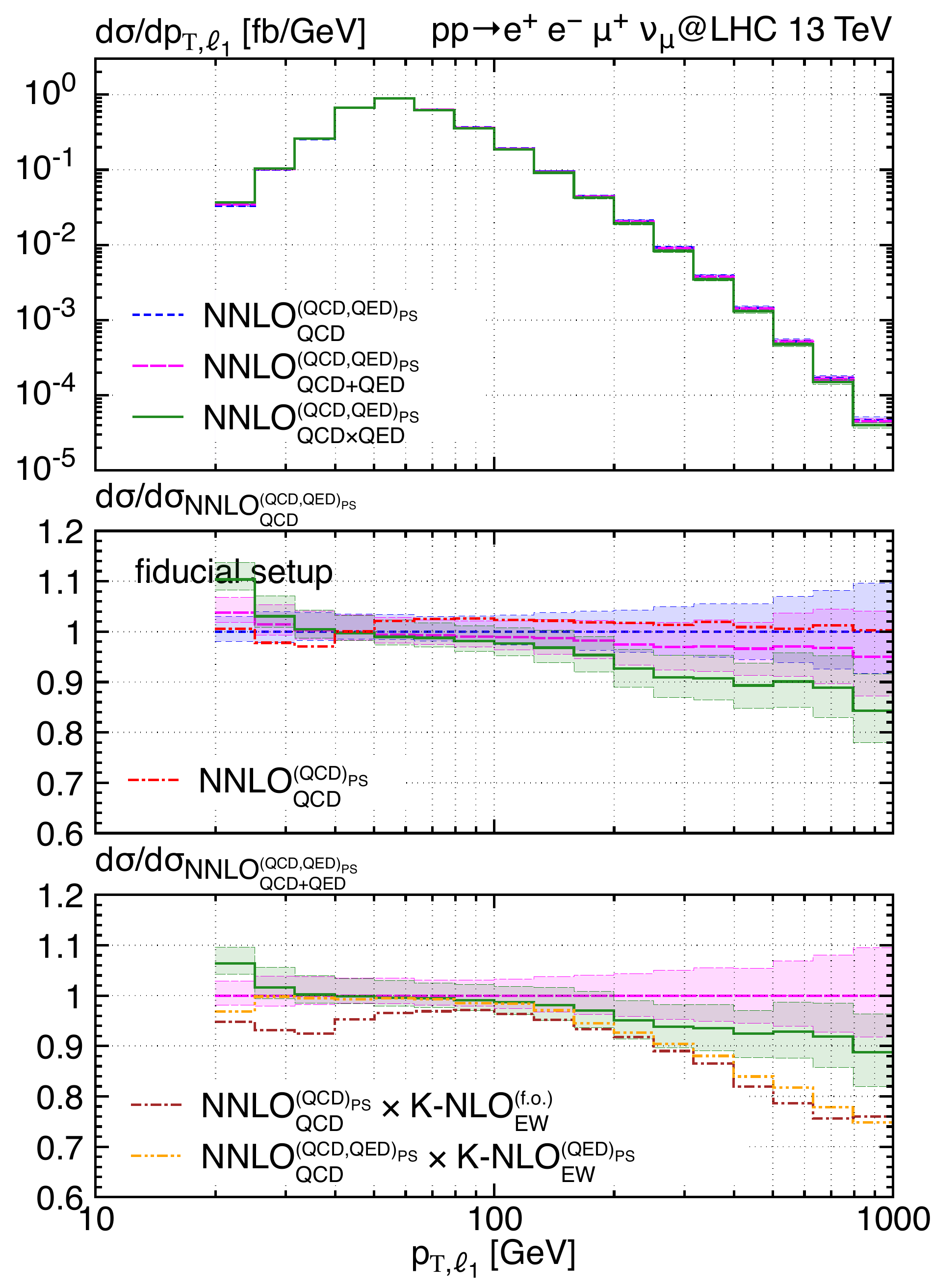} 
\end{tabular}\vspace{-0.15cm}
\begin{tabular}{cc}
\includegraphics[width=.31\textheight]{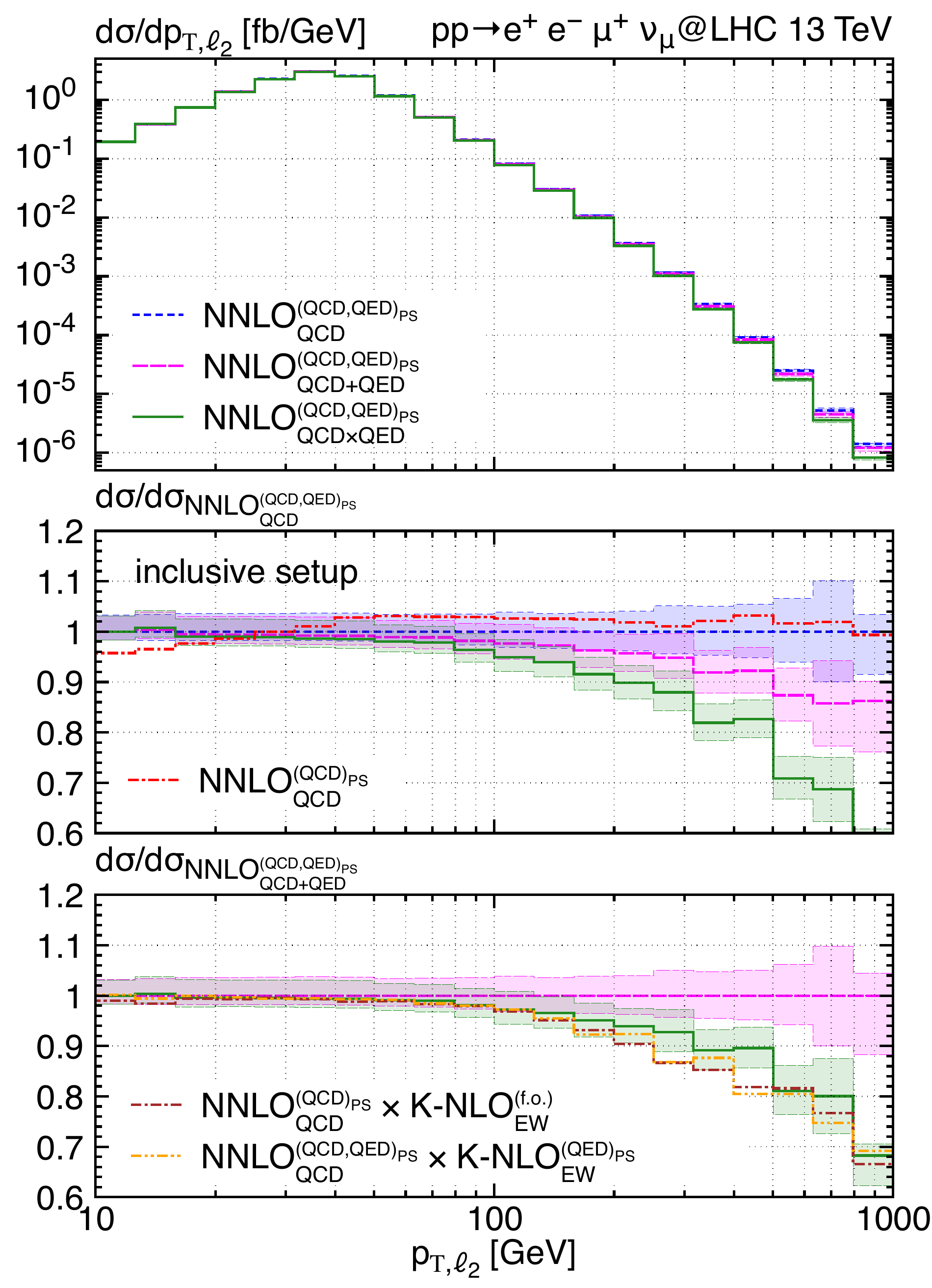} 
&
\includegraphics[width=.31\textheight]{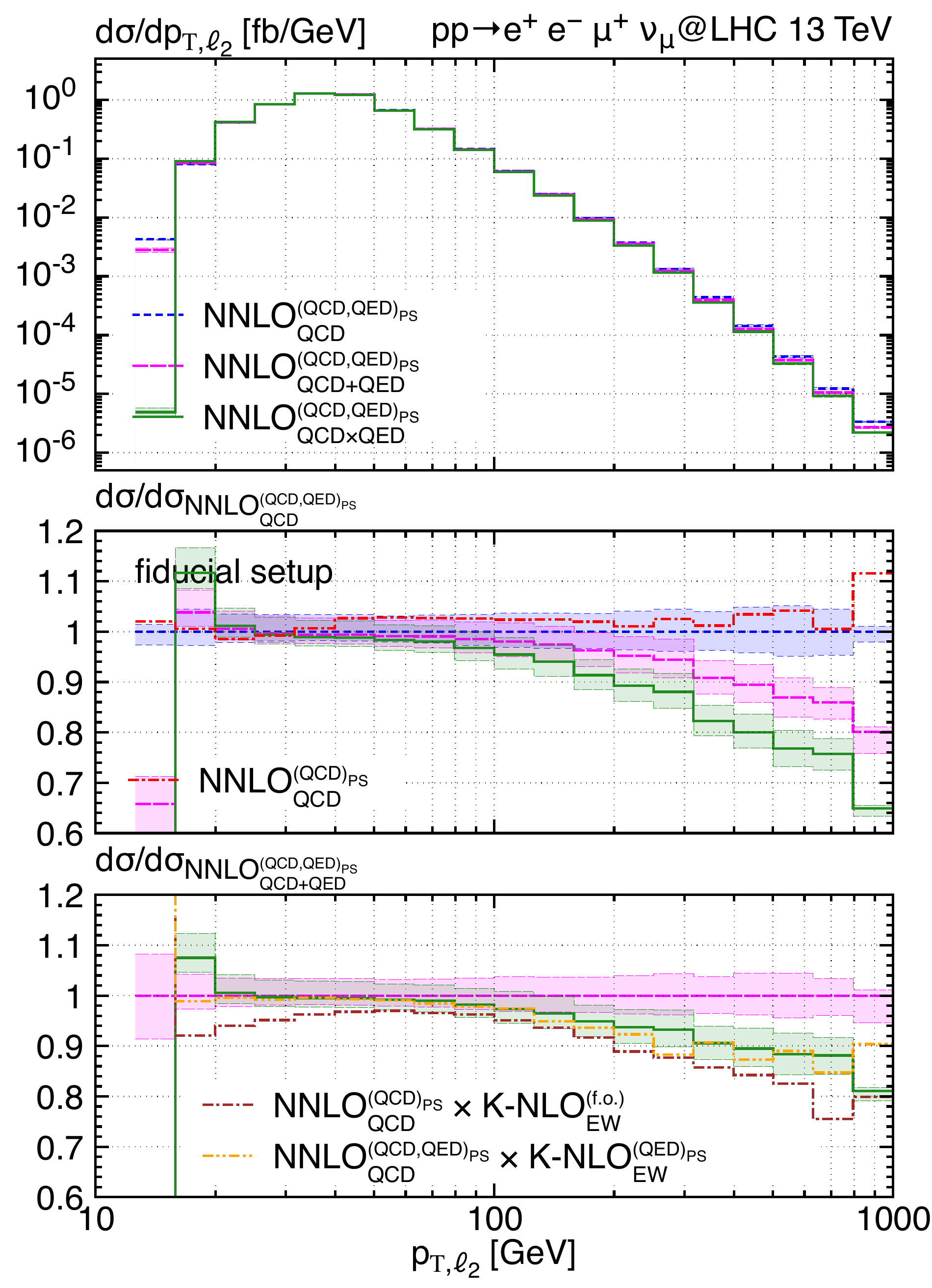}
\end{tabular}
\caption{\label{fig:results_ptl} Differential distributions in the
  leading lepton transverse momentum ($\ptlone$) (top) and subleading
  lepton transverse momentum ($\ptltwo$) in $W^+Z$ production in the
  \setupinclusive{} (left) and \setupfiducial{} (right). See text for
  details.}
\end{center}
\end{figure}

\begin{figure}[t]
\begin{center}\vspace{-0.2cm}
\begin{tabular}{cc}
\includegraphics[width=.31\textheight]{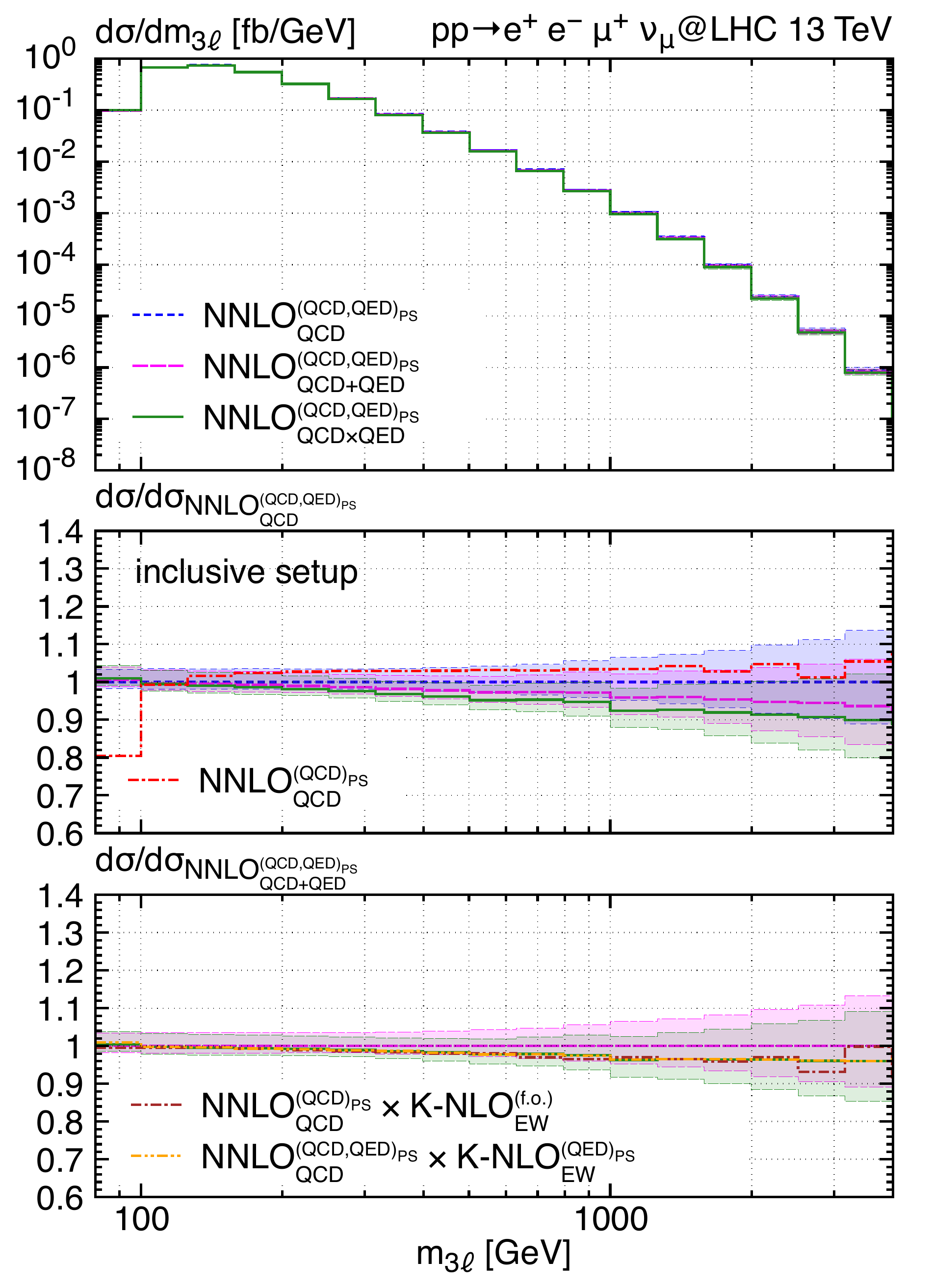} 
&
\includegraphics[width=.31\textheight]{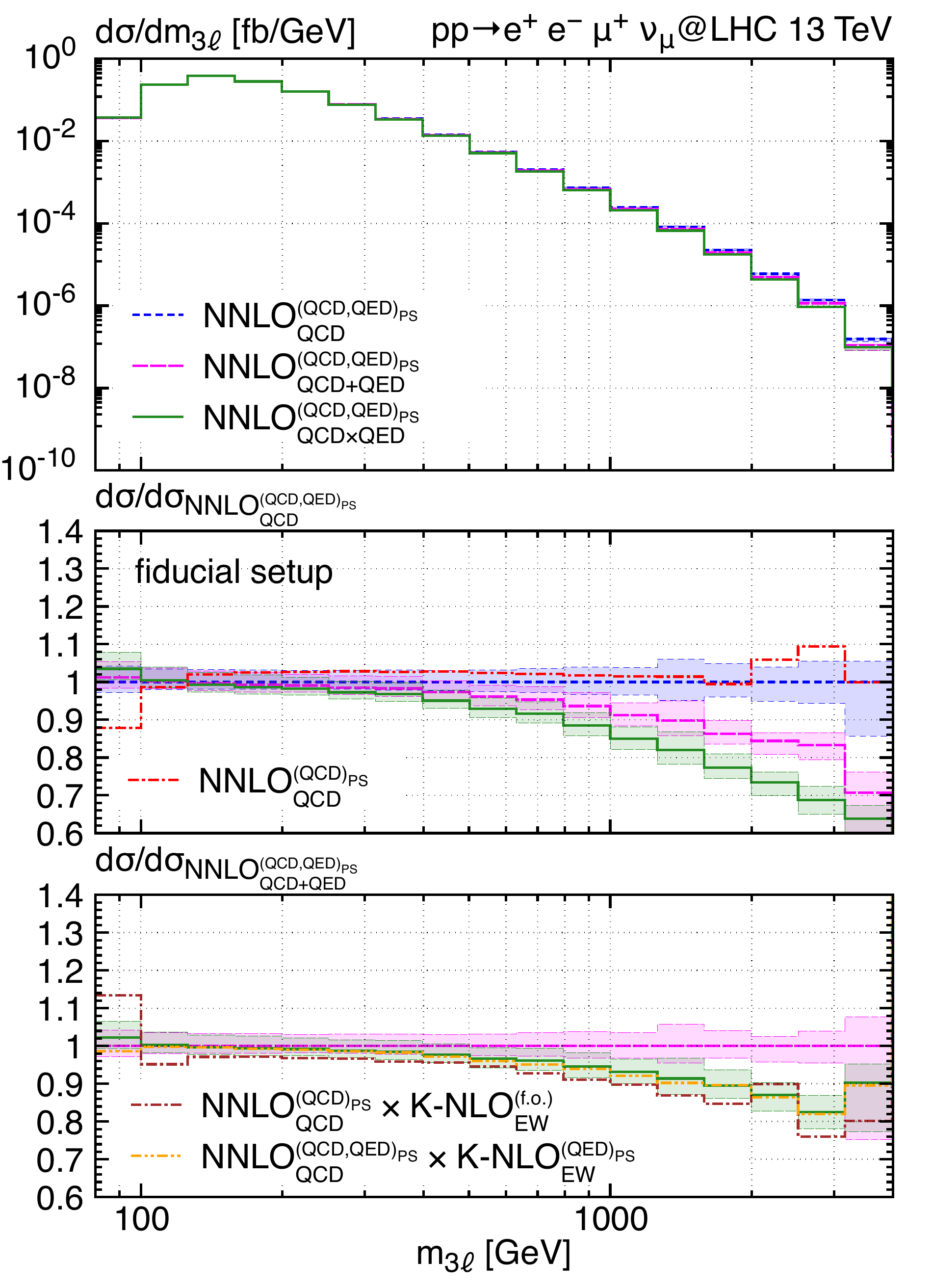}
\end{tabular}
\caption{\label{fig:results_mlll} Differential distributions in the
  invariant mass of the three final-state leptons in $W^+Z$ production
  in the \setupinclusive{} (left) and \setupfiducial{} (right). See
  text for details.}
\end{center}
\end{figure}

For the rest of the discussion we turn to high-energy tails of
differential distributions in
\figs{fig:results_met}--\ref{fig:results_mlll}, which are relevant in
particular for new-physics searches at the energy frontier. For all
observables under consideration we show results in both the
\setupinclusive{} and the \setupfiducial{}, as defined in
\tab{tab:cuts}.  In order to render the high-energy tails visible, in
these figures we employ a logarithmic scale (and binning) on the
$x$-axis. Moreover, we have added the central \multqcdfull{}
predictions (orange, dash-double-dotted) in the lower ratio insets for
comparison.

We start the discussion of high-energy observables with the
distribution in the missing transverse momentum ($\ptmiss$), for which
corresponding plots in the \setupinclusive{}\,(left) and in the
\setupfiducial{}\,(right) are shown in \fig{fig:results_met}.  Here,
by and large, both the \setupinclusive{} and the \setupfiducial{} show
very similar results.  Comparing the \qcdqcd and \qcdfull curves in
the first ratio panel we find percent level effects from QED emissions
in the entire considered $\ptmiss$ range. By contrast, the NLO EW
corrections are significantly enhanced at large $\ptmiss$ values due
to the appearance of EW Sudakov logarithms. Corrections in the
multiplicative combination scheme \QCDtEW{} reach about
$-15\%$ at $\ptmiss=500$~GeV, while in the additive scheme
\QCDpEW{} they reach about $-4\%$ for the same value of
$\ptmiss$. These differences can be explained due to large NLO QCD
corrections plaguing the $\ptmiss$ distribution.  These large QCD
effects are known as 'giant $K$-factors' as discussed in
\sct{sec:combination} and originate from hard vector-boson plus jet
topologies with an additional soft vector boson. As discussed in
\citere{Grazzini:2019jkl} in this situation it is reasonable to
consider the average between \QCDtEW and \QCDpEW as nominal
prediction, and their spread can be interpreted as $\order{\alpha_S
  \alpha}$ uncertainty band.  Looking at the lower ratio inset, we
observe that qualitatively the fixed-order NLO$_{\rm EW}$ $K$-factor
approximation and the parton-shower matched \QCDtEW{} and
\multqcdfull{} predictions follow the same trend.  However, at low
$\ptmiss{}$ one can observe differences of the fixed-order
approximation up to about 5\%, which are present only in the
\setupfiducial{} and thus induced by the fiducial cuts.  In this
regime these differences are in fact larger than the remaining scale
uncertainties, indicating the relevance of using the NLO EW
parton-shower matched computations instead of the fixed-order
approximation.  In the deep tail of the $\ptmiss{}$ distribution, on
the other hand, we see that our nominal \QCDtEW prediction shows a
somewhat smaller EW Sudakov suppression than predicted at fixed order
and by the \multqcdfull{} combination, where the NLO EW $K$-factor is
computed turning on only the QED shower, but without QCD shower.  We
have verified that these differences can be traced back to
aforementioned giant QCD $K$-factor effects, which are generated by
the QCD emissions in the partons-shower matching of the \QCDtEW
combination. Indeed, after applying a suitable (dynamical) veto
against QCD radiation to select configurations where both vector
bosons are sufficiently hard, as suggested in
\citere{Grazzini:2019jkl}, we found the two parton-shower matched
predictions to be practically indistinguishable.

Next in \fig{fig:results_ptl} we turn to the discussion of the
distributions in the transverse momentum of the leading ($\ptlone$)
and the subleading ($\ptltwo$) charged leptons.  These distributions
constitute important experimental observables for both new-physics
searches and for constraining anomalous couplings.  By and large, the
qualitative behaviour of these distributions and the relative
corrections is very similar to the one observed for the $\ptmiss{}$
distribution: NNLO QCD scale uncertainties increase from the level of
few percent in the bulk to about $10\%$ in the tail of the
transverse-momentum distributions, while the EW corrections yield
important shape distortions with increasing negative corrections in
the high transverse-momentum tail. At low transverse momenta, fiducial
cuts induce important effects that can be described appropriately only
by using the parton-shower matched QCD--EW combinations. At high
transverse momenta, our nominal multiplicative \QCDtEW{} combination
features smaller NLO EW corrections than predicted at fixed order and
the \multqcdfull{} result. We recall that this difference is induced
by giant QCD corrections created by the QCD shower in the \QCDtEW{}
combination, which affect the NLO EW $K$-factor. Indeed, for the
$\ptltwo$ distribution, which is less affected by giant $K$-factor
effects, the three multiplicative predictions in the lower ratio panel
are much closer to one another than in the case of the $\ptlone$
distribution.  At variance with the $\ptmiss{}$ distribution, we
observe non-trivial QED effects at low transverse momenta. These
effects are visible both comparing the \qcdqcd and \qcdfull curves in
the central ratio panel and when comparing \QCDtEW with the
fixed-order approximation in the lower ratio panel.

Finally, in \fig{fig:results_mlll} we consider the invariant-mass
distributions of the three charged final-state leptons (\mlll), which
can be seen as a proxy for the (unobservable) invariant mass of the
full \wz{} system.  Also for this observable the EW corrections are
negative and increase in the high-energy tail due to the appearance of
EW Sudakov logarithms.  It is interesting to notice that the EW
corrections substantially increase as soon as fiducial cuts are
applied. Indeed, when moving from the \setupinclusive{} to the
\setupfiducial{}, negative EW effects at invariant-mass values of
$\sim 2$\,TeV increase from about $-10$\% to about
\mbox{$-20$--$30$\%}, rendering the inclusion of EW corrections
crucial in such high-energy phase-space regions. The origin of this
effect can be explained as follows: In the \setupinclusive{}, the
high-$\mlll$ region is populated by very forward leptons at large
rapidities. In this regime, not all Mandelstam invariants $s_{ij}$ are
large, resulting in a suppression of the double Sudakov logarithms
$\ln^2(|\hat s_{ij}|/M_W^2)$.  In the \setupfiducial{}, the very
forward regime of the leptons is removed by their rapidity
requirements, resulting in the expected EW Sudakov enhancement.  We
also point out that the $\mlll$ distribution is not plagued by giant
QCD $K$-factor effects. Therefore, the \QCDtEW and the \multqcdfull{}
parton-shower combinations, as well as the fixed-order approximation
of the EW corrections, practically coincide with one another in the
$\mlll$ tail.

We note that in our numerical results presented here we have not
included any results for the \addqcdfull{} combination, because we
have not found any meaningful differences with respect to the
\QCDpEW{} curves shown in the presented figures, and we consider both
approaches equally appropriate as an additive combination.  Similarly,
the \multqedfull{} combination yields very similar results to the
\QCDtEW{} predictions, including the sensitivity to giant QCD
$K$-factors generated by the QCD shower for certain observables (which
are again alleviated through a proper veto against QCD
radiation). Consequently, we also refrain from showing any
\multqcdfull{} results.

\subsection{Comparison to data}
\label{sec:comparisondata}

\begin{figure}[t]
\begin{center}\vspace{-0.2cm}
\begin{tabular}{cc}
\includegraphics[width=.31\textheight]{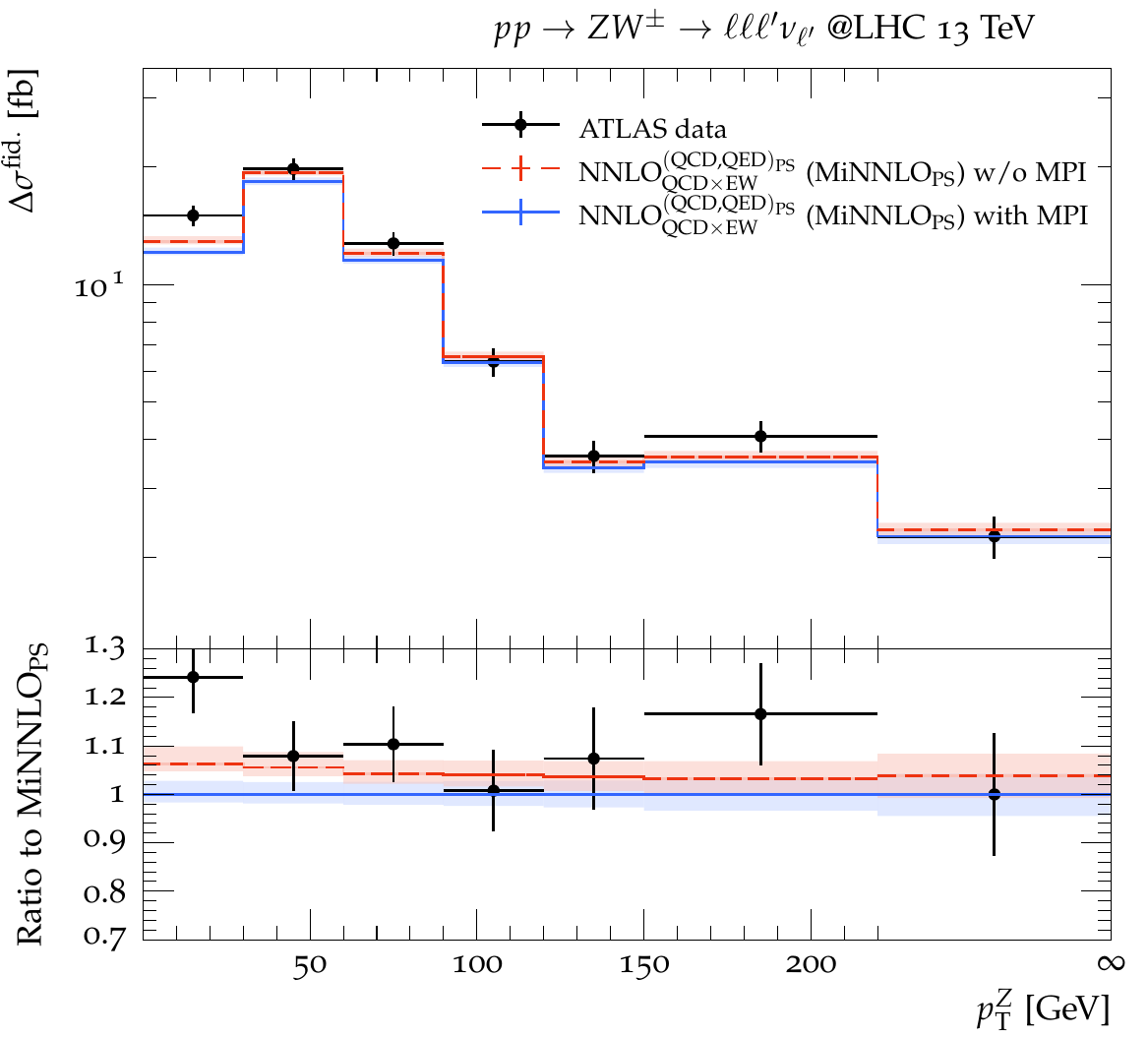}
&
\includegraphics[width=.31\textheight]{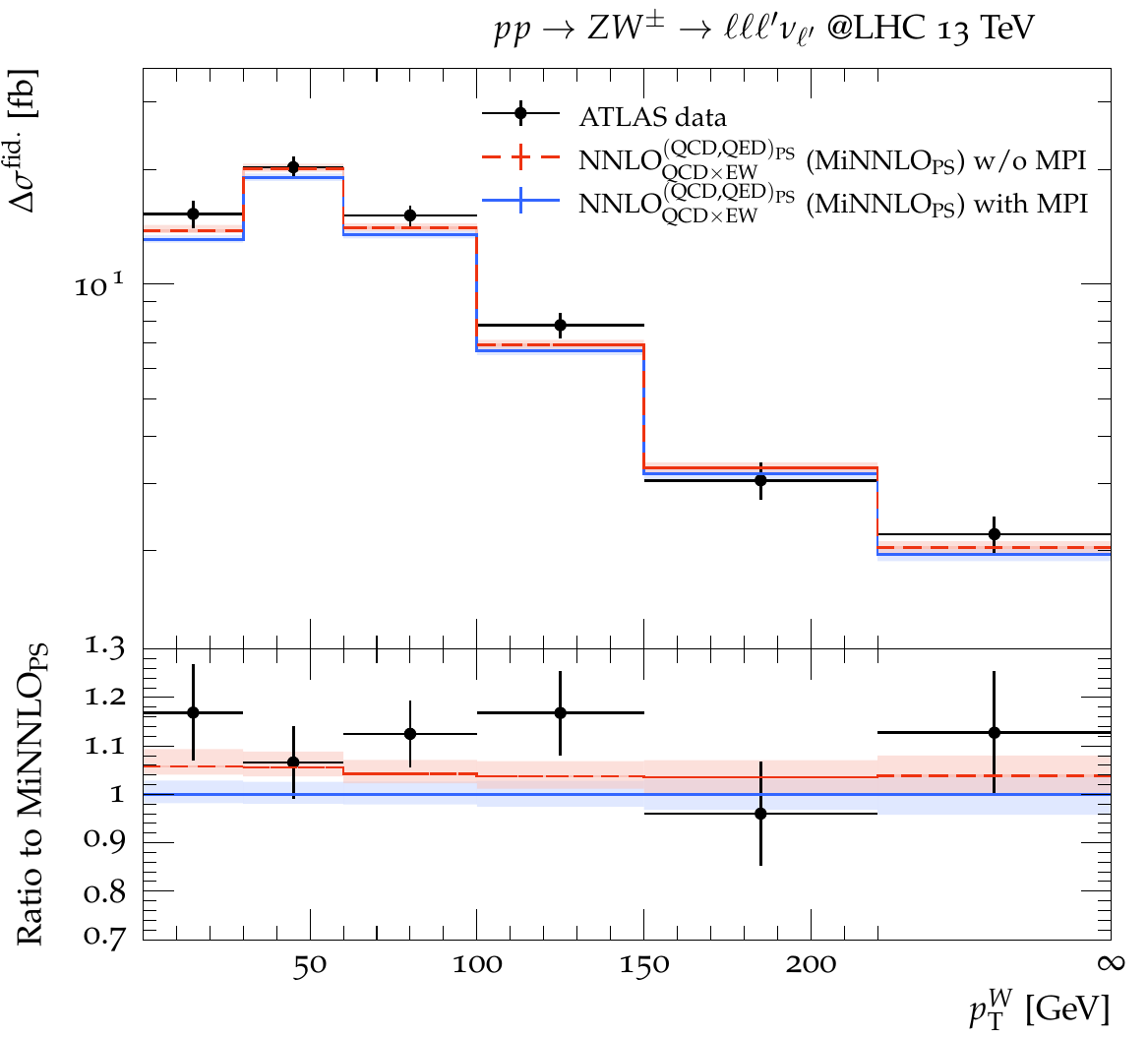}
\end{tabular}\vspace{0.3cm}
\begin{tabular}{cc}
\includegraphics[width=.31\textheight]{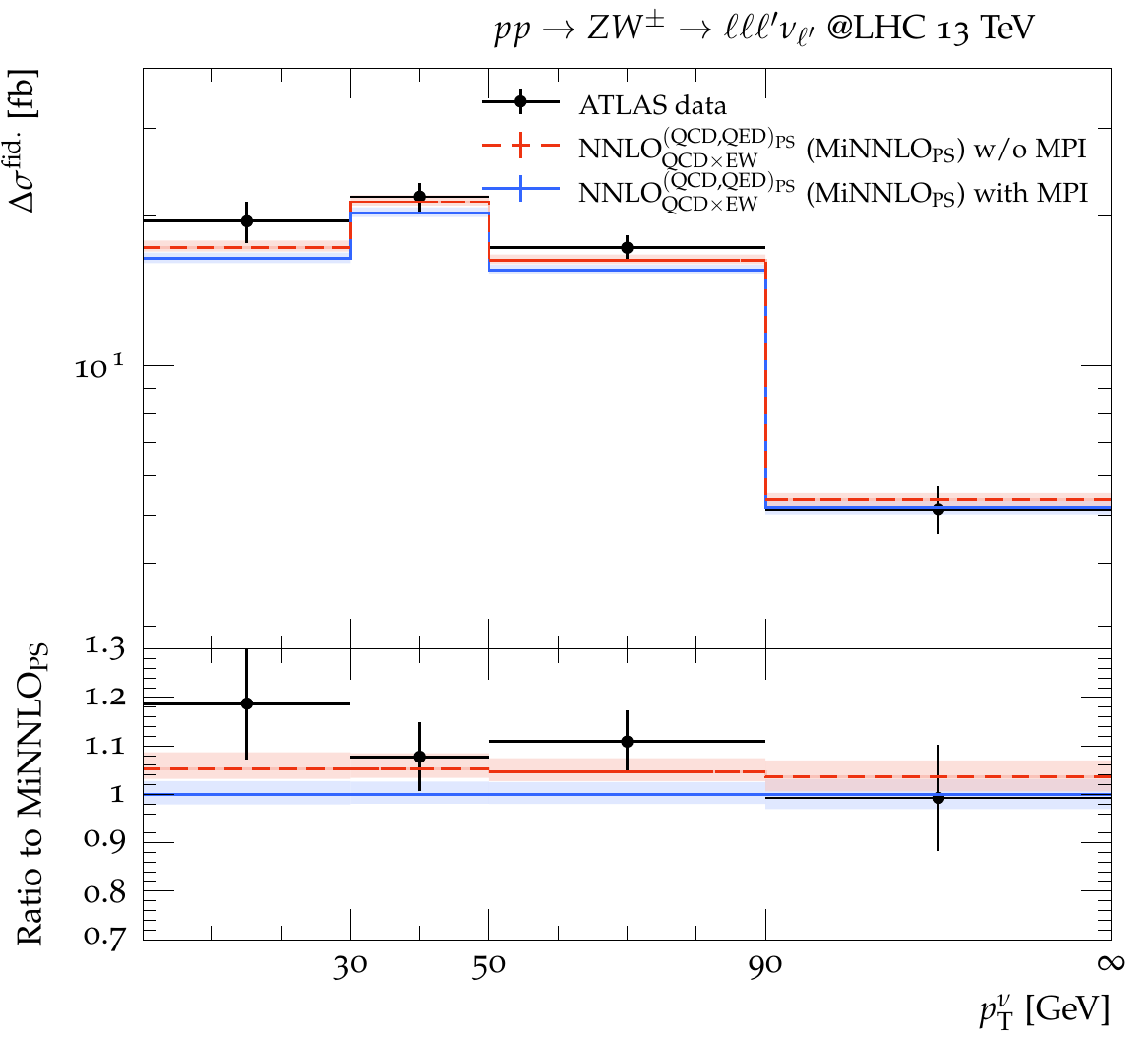}
&
\includegraphics[width=.31\textheight]{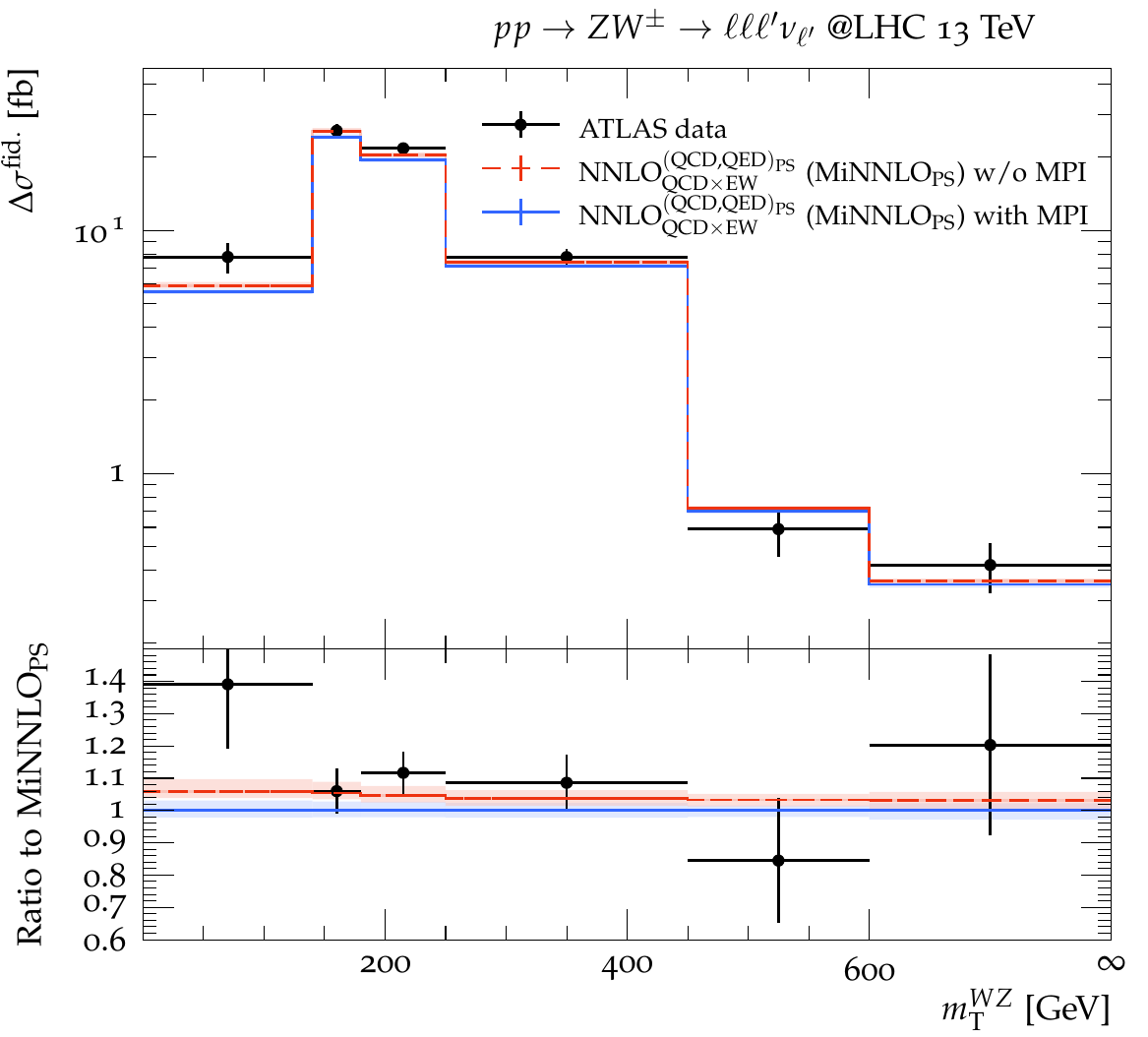}
\end{tabular}\vspace{0.3cm}
\begin{tabular}{cc}
\includegraphics[width=.31\textheight]{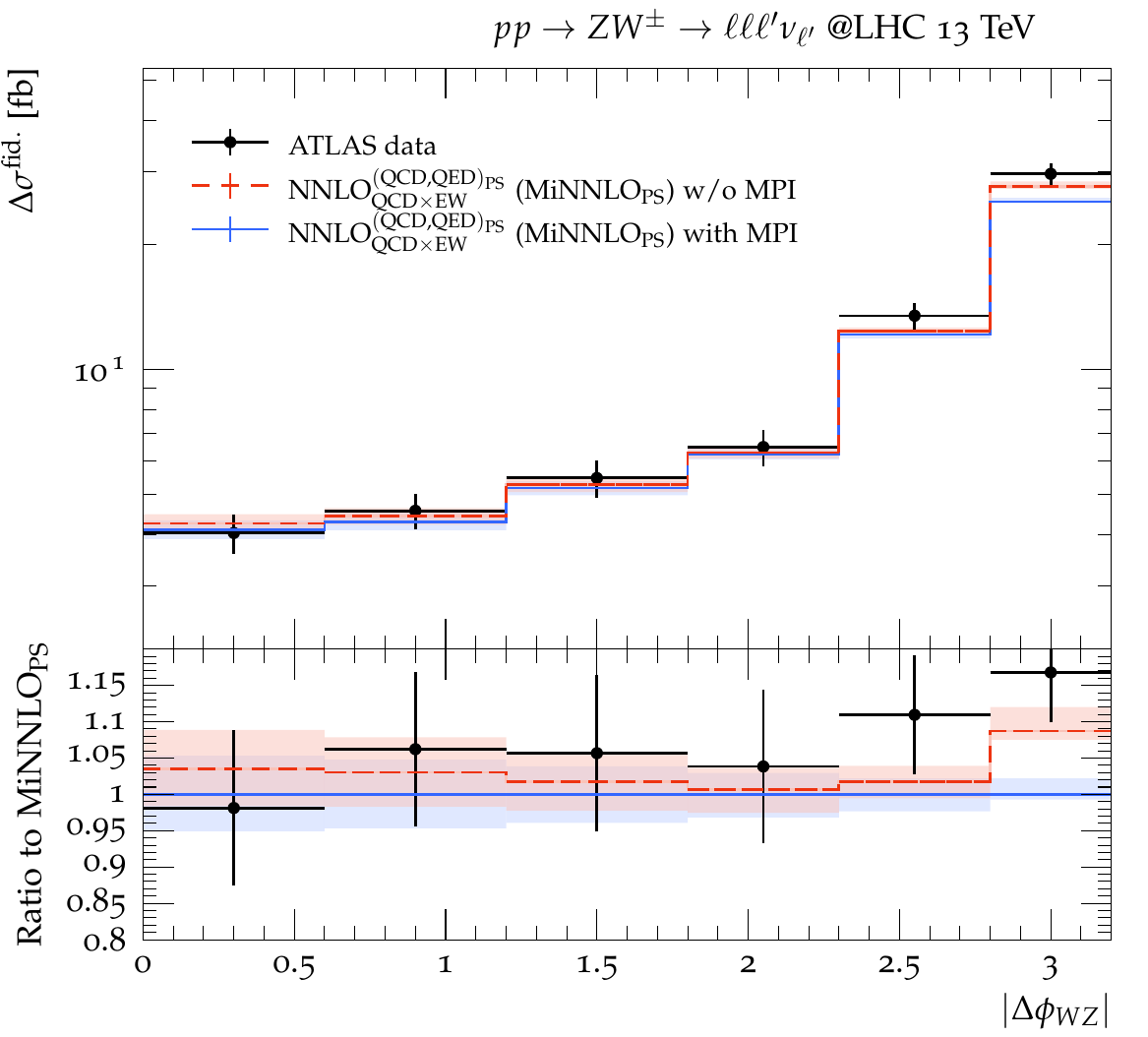}
&
\includegraphics[width=.31\textheight]{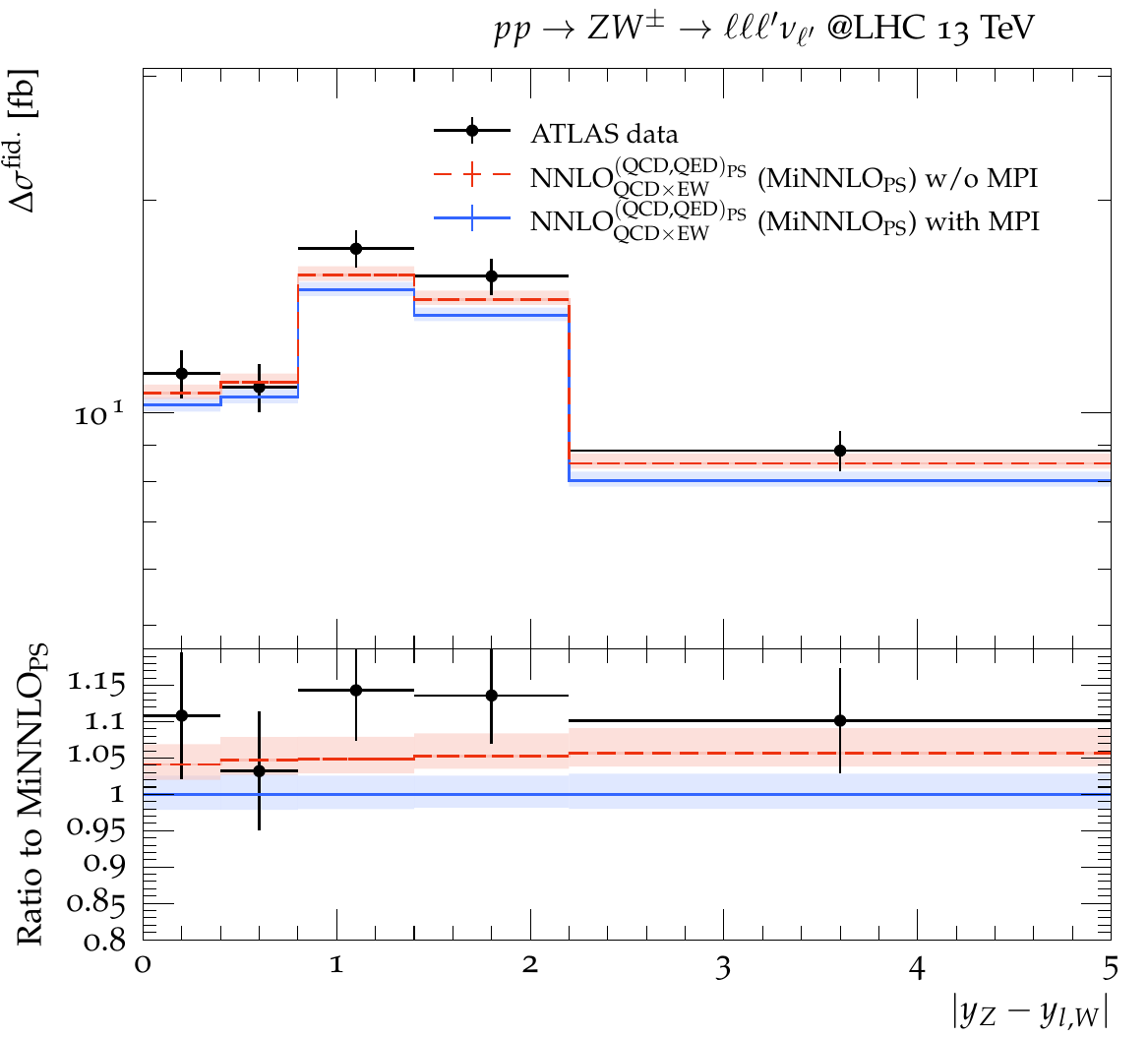}
\end{tabular}
\caption{\label{fig:data1} Comparison of our default \minnlo{} prediction \QCDtEW with MPI effects (blue, solid) and without (red, dashed) against the ATLAS data from the analysis~\cite{ATLAS:2019bsc}.}
\end{center}
\end{figure}

Finally, we compare our default \minnlo{} predictions, which
correspond the multiplicative combination of NNLO QCD and NLO EW
corrections, defined as ${\rm NNLO}_{\rm QCD\times EW}^{\rm (QCD,
  QED)_{\rm PS}}$ in \eqn{eq:best}, to a recent differential
measurement of \wz{} production performed by the \mbox{ATLAS}
collaboration \cite{ATLAS:2019bsc}. The fiducial phase space is defined in \sct{sec:settings}.
To facilitate this comparison, we employ the corresponding {\sc Rivet}
routines provided on the {\tt HEPdata}
webpage\footnote{\href{https://www.hepdata.net/record/ins1720438}{https://www.hepdata.net/record/ins1720438}}
for the analysis of \citere{ATLAS:2019bsc}.  The differential
distributions measured in \citere{ATLAS:2019bsc} correspond to the
averaged \wz{} cross section of all combinations of electrons and
muons in the final state. Thus, the \wz{} cross section for the
same-flavour channels ($e^+e^-e^\pm\nu_e$, $\mu^+\mu^-\mu^\pm\nu_\mu$)
and different-flavour channels ($e^+e^-\mu^\pm\nu_\mu$,
$e^+e^-\mu^\pm\nu_\mu$) are summed and divided by four, i.e.\ by the
total number of channels.  For simplicity, our \minnlo{} predictions
rely on the different-flavour channel only, in which case the $W$ and
the $Z$ boson can be unambiguously reconstructed.\footnote{We stress
  again that the codes developed in this work can simulate both
  different-flavour and same-flavour events.} Indeed, using the
different-flavour result for the same-flavour channel has been shown
to be an excellent approximation~\cite{Grazzini:2017ckn}, especially
when the vector bosons are reconstructed according to the
``resonant-shape'' identification procedure, which yields a very high
performance in reliably reconstructing the $W$ and $Z$ bosons. The
ATLAS analysis of \citere{ATLAS:2019bsc} that is considered here is
based exactly on this kind of reconstruction.

In \fig{fig:data1} we show the distributions in the transverse
momentum of the reconstructed $Z$ boson ($\ptz{}$), of the
reconstructed $W$ boson ($\ptw$) and of the neutrino ($\ptnu$),
i.e.\ the missing transverse momentum, the transverse mass of the
\wz{} system ($\mtwz$), defined as
 \begin{align}
 \mtwz =\sqrt{ \left(\sum_{i=1}^{4}p_{\text{\scalefont{0.77}T,$i$}}\right)^2- p_{\text{\scalefont{0.77}T,$WZ$}}^2} \,,
 \end{align}
 where the sum runs over the three charged leptons and the neutrino,
 as well as the difference in the azimuthal angle between the
 reconstructed $Z$ and $W$ bosons ($\dphiwz$) and the absolute
 rapidity difference between the reconstructed $Z$ boson and the
 charged lepton coming from the $W$ decay ($\dyZlW$).  The main frame
 of all plots in \fig{fig:data1} shows the absolute cross section per
 bin for the data and our default prediction, while the lower panel
 shows the ratio to \minnlo{}. The last bin in all kinematically
 unbounded distributions shall be understood as an overflow bin, which
 is indicated by the infinity symbol on the $x$ axis of its right
 edge.  We note that, except for $\dphiwz$ (where $\dphiwz=\pi$ at
 LO), all distributions are defined already at LO in the Born phase
 space and are therefore genuinely NNLO QCD and NLO EW accurate.  In
 these distributions, our \minnlo{} predictions provide a remarkable
 agreement with data. This is true both in the bulk region of the
 cross section, where NNLO QCD corrections are vital, and in the tails
 of the distributions, where NLO EW corrections become relevant. 
 It is interesting to notice, however, that MPI effects are relatively large, 
 lowering the predictions by about $5\%$ and, overall, leading to a slightly 
 worse agreement with data.
 With remaining scale uncertainties of only a few percent, the theory
 predictions can be regarded as extremely precise.  On the contrary,
 the experimental errors have not reached (yet) a comparable level of
 accuracy with the given statistics of $36.1$\,fb$^{-1}$, which can
 nevertheless be expected to change once the full Run\,2 data is
 considered and, even more so, once Run\,3 data becomes available in
 the future.

As anticipated, the $\dphiwz$ distribution is filled for $\dphiwz<\pi$
only upon inclusion of higher-order corrections, since this region
requires a recoil for the \wz{} system, for instance by a jet. As a
result, this distribution is effectively only NLO QCD accurate at low
$\dphiwz$, which is also reflected by the slightly enlarged
scale-uncertainty band.  Moreover, around the $\dphiwz=\pi$ threshold
the distribution becomes sensitive to soft-gluon effects, which are
accounted for by the matching to the parton shower. Despite these
caveats, the agreement between the predicted and measured $\dphiwz$
distribution is excellent and within at most one standard deviation
for all bins.

\section{Conclusions}
\label{sec:summary}
In this paper we have presented the first calculation of NNLO QCD
corrections matched to parton showers for \wz{} production at the
LHC. We consistently combined this calculation with NLO EW corrections
that we also matched to parton showers. This is the first time such
accuracy is reached in the simulation of a LHC process.  This is
achieved via two separate implementations within the \POWHEGBOXRES{}
code, consisting of one \minnlo{} \wz{} generator and one \POWHEG{}
generator for \wz{} production including both NLO QCD and NLO EW
corrections.

Our results have been validated against fixed-order calculations in
NNLO QCD and NLO EW obtained from \Matrix{}. For observables related
to the colour-singlet final state, which are inclusive over radiation
and not sensitive to soft-gluon effects, our \minnlo{} predictions are
fully compatible with fixed-order NNLO QCD predictions within scale
uncertainties. As far as our \POWHEG{} NLO EW implementation is
concerned, the NLO EW cross sections have been shown to be essentially
identical in the relevant phase space regions to their fixed-order
counterparts obtained with \Matrix+\OpenLoops{}.

Phenomenological results have been discussed in detail for various
combinations of NNLO QCD and NLO EW predictions. In particular, we
have investigated different additive and multiplicative combination
approaches. In these combinations we consistently included and/or
excluded QCD and QED emissions in the shower matching in certain parts
of our combined NNLO QCD and NLO EW accurate calculation, avoiding any
double counting while preserving the desired accuracy.  We have
considered a number of distributions where different physical effects
become relevant, both in the fully inclusive phase space and in
presence of fiducial cuts.

We find that QED effects are crucial for observables related to the
dressed charged leptons due to recoil effects, which can not be fully
reabsorbed by the dressing. This is particularly important in the
line-shape distribution of the reconstructed $Z$ boson, but also below
the peak in other invariant mass distributions as well as at small
transverse momenta of the charged leptons. We observed that in most
cases the full tower of QED emissions as generated by the QED shower
is well approximated when considering only the first photon emission
through a fixed-order calculation.  However, in certain cases
differences up to 10\% remain.

In the bulk of the cross sections (away from high-energy tails), we
find that just accounting for EW effects through a QED shower applied
only to the NNLO QCD \minnlo{} predictions already serves as a very
good approximation.  Instead, full NLO EW corrections are most
important in the high-energy tails of distributions, where they lead
to a strong suppression of the cross section due to large EW Sudakov
logarithms.  We argued that, generally speaking, a multiplicative
scheme is more suitable for describing high-energy regions of phase
space. Indeed, since EW Sudakov logarithms are expected to largely
factorize with respect to QCD corrections, a multiplicative scheme
provides a better approximation of the missing mixed QCD--EW
contributions.  However, this general assumption does not hold for
distributions plagued by giant QCD $K$-factors, which are dominated by
\wz+jet topologies.  In fact, in such a situation also the computation
of an NLO EW $K$-factor in a parton-shower matched calculation can be
affected by whether QCD shower emissions are taken into account or
not, as those emissions can create giant QCD correction effects in the
high-energy tails.  However, by selecting phase-space configurations
involving two hard vector bosons, e.g.\ through appropriate
(dynamical) jet-veto prescriptions, any giant $K$-factor issues can be
avoided.  Moreover, we have shown that in high-energy tails EW
corrections obtained at NLO+PS can be well approximated by using a NLO
EW $K$-factor computed at fixed order.  This justifies the application
of NLO EW $K$-factors on QCD predictions, which have been widely used
by the LHC collaborations.  However, when applying these factors on
showered predictions, QED shower effects have to be turned off in
order to avoid a double counting of QED radiation.  Nevertheless, we
also observed that fiducial cuts can affect the size of NLO EW
corrections in the bulk region of the cross section, rendering
parton-shower matched NLO EW predictions important for precision
measurements.

Not least, we have employed our default predictions including NNLO QCD
and NLO EW corrections in a direct comparison to recent \mbox{ATLAS}
data at 13\,TeV. We find remarkable agreement between theory
predictions and experimental data for all observables. At the moment,
the experimental accuracy is still limited by statistical
uncertainties for the considered $36.1$\,fb$^{-1}$ analysis. However,
with scale uncertainties of only few percent our \minnlo{} predictions
will facilitate high-precision studies for \wz{} production in the
future.  The Monte Carlo generators developed in this work will be
made publicly available within the \POWHEGBOXRES{} framework.

Finally, in order to control remaining QCD--EW mixed uncertainties in
observables subject to giant $K$-factors, and in order to directly
produce NNLO QCD and NLO EW accurate events, instead of combining such
predictions a posteriori at the level of distributions, we will
consider suitable extensions of the \minnlo{} method in the
future. This will allow to simulate fully differential events for
\wz{} production and other colour-singlet processes that are NNLO QCD
and NLO EW accurate.

\section*{Acknowledgements}

We would like to thank Christian G\"utschow for clarifications
regarding the {\sc Rivet} implementation of the recent ATLAS \wz{}
analysis.  We also would like to thank Javier Mazzitelli for his help
with issues related to {\sc Rivet}.  D.L.\ and M.W.\ would like to thank
CERN, where some of this work has been pursued, for their kind
hospitality.  M.W.\ also thanks DESY for kind hospitality in the
context of the ``Theorist of the Month'' programme to strengthen the
exchange between theorists and experimentalists.  J.M.L.\ is supported
by the Science and Technology Research Council (STFC) under the
Consolidated Grant ST/T00102X/1 and the STFC Ernest Rutherford
Fellowship ST/S005048/1. S.Z.\ is supported by the International Max
Planck Research School (IMPRS) on ``Elementary Particle
Physics''. D.L., M.W, G.Z, and S.Z.\ acknowledge MIAPP under the
program "Gearing up for High-Precision LHC Physics'' for hospitality
while this work was being finalized.

\bibliography{MiNNLO}
\bibliographystyle{JHEP}

\end{document}